\documentclass[final,5p,times,twocolumn]{elsarticle}

\usepackage{epstopdf}
\usepackage{epsfig}
\usepackage{graphicx}
\usepackage{subfigure}
\usepackage{amsmath}
\usepackage{lineno}
\usepackage[tableposition = top]{caption} % captions
\usepackage{booktabs}                     % horizontal lines in tables
\usepackage[per-mode = symbol]{siunitx}   % physical quantities and SI units
\usepackage{hyperref}

\biboptions{sort&compress}

\journal{Journal of Molecular liquids}

\begin{document}

\begin{frontmatter}

\title{The impact of magnetic field on the conformations of supracolloidal polymer-like structures with super-paramagnetic monomers}
\author[label1]{Mostarac D.}
\author[label2]{Novak E.V.}
\author[label2]{S\'anchez P.A.}
\author[label1,label2]{Kantorovich S.S.}
\address[label1]{University of Vienna, Sensengasse 8, 1090 Vienna, Austria}
\address[label2]{Ural Federal University, Lenin Av. 51, Ekaterinburg, 620000, Russia}

\begin{abstract}
 We investigate the properties of magnetic supracolliodal polymers -- magnetic filaments (MFs) -- with super-paramagnetic monomers, with and without Van der Waals (VdW) attraction between them. We employ molecular dynamics (MD) simulations to elucidate the impact of crosslinking mechanism on the structural and magnetic response of MFs to an applied external homogeneous magnetic field. We consider two models: plain crosslinking, which results in a flexible backbone; and constrained crosslinking, which provides significant stiffens against bending. We find that for plain crosslinking, even a slight increase of the central attraction leads to collapsed MF conformations. Structures that initially exhibit spherical symmetry evolve into cylindrically symmetric ones, with growing magnetic field strength. Plain crosslinking also allows for conformational bistability. MFs with constrained crosslinking tend to, instead, unravel in field. In both crosslinking scenarios, central attraction is able to hinder low-field magnetic response of MFs, albeit the bistability of plainly crosslinked MFs manifests itself also in the high-field region. 
\end{abstract}


\end{frontmatter}

\section{Introduction}

The field of soft matter and the idea of smart, soft matter materials has come a long way since magnetic fluids were first synthesised.\cite{resler1964magnetocaloric} Such materials, with responsiveness to magnetic fields can be made by combining magnetic micro- and nanoparticles (MNPs) with conventional soft materials, such as fluids or gels. Over the years, this idea has grown into a large number of synthetic soft matter systems.\cite{2009-odenbach,zrinyi1998kinetics,weeber2018polymer,volkova2017motion,frank1993voltage,zrinyi1998kinetics,Weeber_2012} Out of these systems,  magnetic filaments (MF),\cite{Dreyfus_2005,2008-benkoski} first synthesized as micron-sized magnetic-filled paramagnetic latex beads forming chains,\cite{1998-furst,1999-furst} open up a plethora of new potential applications.\cite{WANG_2011,wang2014multifunctional,cebers2016flexible,cai2018fluidic} They have been experimentally investigated for artificial swimmer uses,\cite{2005-dreyfus,2008-erglis-mh} cellular engineering,\cite{fayol2013use,gerbal2015refined} and biomimetic cilia designs.\cite{evans2007magnetically,erglis2011three} 

Even though synthesis techniques of MFs are nowadays rather diverse and powerful,\cite{2003-goubault,2005-cohen-tannoudji,2005-singh-lm,2005-singh-nl,2007-martinez-pedrero,2007-evans,2008-benkoski,2009-zhou,2011-benkoski,2011-wang,2012-sarkar,2012-breidenich,busseron2013supramolecular,2014-byrom,hill2014colloidal,2015-bannwarth} there is no clear recipe how to create a system of polymer-like supracolloidal chains with polymer flexibility, whose magnetic response is provided by monomers -- magnetic nanoparticles. Recent advances in programmable DNA-MNP assembly techniques have shown a potential to produce flexible, nanoscale MFs, with a highly controllable micro-structure.\cite{Maye_2009, Sun_2012, Sun_2013, Zhang_2013,Srivastava_2014, Tian_2015,liu2016self} As it stands, however, such filaments have not been achieved and as synthesis attempts have suggested, Van der Waals forces, together with the super-paramagnetic nature of magnetic colloids, might be of high importance. 

Theoretically, MFs have mostly been explored in bulk,\cite{2003-cebers,2004-shcherbakov,2004-cebers,2005-cebers,belovs2006nonlinear,cebers2007magnetic,_rglis_2008,kuznetsov2019equilibrium} and their behaviour when exposed to an external magnetic filed has also been scrutinized.\cite{huang2016buckling,zhao2018nonlinear,wei2016assembly,kuei2017strings,dempster2017contractile,vazquez2017paramagnetic} Theoretical work has been done on designs for artificial swimmers,\cite{2006-gauger,roper2006dynamics,roper2008magnetic} bio-medical applications,\cite{philippova2011magnetic,pak2011high,Saanchez_2015}, micro-mixers,\cite{biswal2004micromixing}, as well as designs for cargo capture and transport. \cite{yang2017magnetic} Magnetic filaments or fibers in micron-scale are valuable for tuning the effective viscosity of magnetorheological suspensions. \cite{lopez2009magnetorheology}
To the best of our knowledge, there are currently no comparative studies of nanoscale, super-paramagnetic MFs, where the colloids can also interact via a central attraction, akin to having Van der Waals forces present in the system, while exposed to an external magnetic field. Furthermore, putting the properties of such systems in the context of diverse crosslinking scenarios has not been done as of yet and remains a tantalising question. 
In this manuscript we employ molecular dynamics (MD) simulations to fill the aforementioned gap in understanding of fundamental properties of magnetic filaments. We make two crosslinking models for super-paramagnetic MFs, which allows us to explore the effects of dipole interactions, magnetisation and central forces in the context of either a rather flexible filament backbone or a backbone that has significant stiffens against bending. We present a comparative analysis of their equilibrium, structural and magnetic properties, in constant, homogeneous magnetic fields. 

The paper is structured as follows: in Section \ref{sec-model} we present the details of our coarse-grained modelling approach, magnetic properties of MNPs, magnetic and Van der Waal interactions, modelling of super-paramagnetic MNPs, simulation protocol as well as a description of reduced units. We introduce two distinct crosslinking approaches for super-paramagnetic MFs. We proceed to discuss our Results in Section \ref{sec-res}. We present how a choice of crosslinking approach, strength of central attraction and/or the magnetic nature of the colloids, affects the structural and magnetic properties of filaments with super-paramagnetic colloids. In Section \ref{sec-conc}, we provide a short summary and prospects of our study.

\section{Model and simulation details}\label{sec-model}
We consider MFs as consisting of organised, mono-disperse, magnetic $N=20$ monomers, modeled as identical super-paramagnetic spherical particles with a characteristic reduced diameter $\sigma = 1$ and reduced mass $m=1$, carrying point magnetic dipole moments located at their centers, denoted by $\vec{\mu}$. The length of 20 monomers is a good compromise, where the computational resources necessary for the study are kept reasonable, while MFs remains long enough to reflect polymeric properties.
The steric repulsion between the colloids is modelled via the Weeks-Chandler-Andersen (WCA) pair potential\cite{weeks1971role}:
\begin{equation}
U_{WCA}(r)=
\begin{cases}
U_{LJ}(r)-U_{LJ}(r_{cut}),\hspace{40pt} r<r_{cut}\\
\hspace{30pt} 0, \hspace{78pt} r\geq r_{cut}
\end{cases}
\label{eq:wca}
\end{equation}
where $U_{LJ}(r)$ denotes the Lennard-Jones (LJ) potential:
\begin{equation}
U_{LJ}(r)=4\epsilon \Biggl\{(\sigma/r)^{12}-(\sigma/r)^6\Biggr\}
\label{eq:LJ}
\end{equation}
and the value of the cutoff is $r_{cut}=2^{1/6}\sigma$. Parameter $\epsilon=1$ defines the depth of the LJ-potential and how the strength of the inter-particle repulsion changes when their center-to-center distance $r$ decreases. As suggested in the introduction, we explore two distinct crosslinking mechanisms between the colloids within a filament, which we will henceforth refer to plan and constrained crosslinking, respectively. By plain crosslinking, it is to be understood that neighbouring colloids of an MF are bonded center-to-center via the $FENE$ potential, given by:
\begin{equation}
U_{FENE}(r)=\dfrac{-K_f r^2_f}{2}\ln\Biggl\{1-\Bigl(\dfrac{r}{r_f}\Bigr)^2\Biggr\},
\label{eq_fene}
\end{equation}
where $r_f=2\sigma$ and $K_f=2.5$ are the maximum extension and the rigidity of a FENE bond respectively. Modelling the effects of crosslinking in this way ensures close contact of the colloids without restricting their rotations. Here, the head-to-tail arrangement of the dipole moments, will be achieved through the cooperative influence of the magnetic dipole field and the external magnetic field (if applied), exclusively. By constrained crosslinking, it is to be understood that, in addition to the center-to-center $FENE$ potential between the colloids, we add an isotropic bonding pair potential between first-nearest neighbors:
\begin{equation}
U_{bend}(\phi)=\dfrac{K_b}{2}(\phi-\phi_0)^2,
\label{eq_bend}
\end{equation}
where $\phi$ is the angle between the vectors spanning from particle $i$ to its nearest neighbour particle pair $(i-1,i+1)$, $i \in [2,N-1]$. $K_b=3.2$ is the bending constant, while $\phi_0=\pi$ is the equilibrium bond angle, both chosen so that out of field end-to-end distance between crosslinking models matches as closely as reasonably possible. The expression in Eq. \eqref{eq_bend} is a harmonic angle dependent potential. In this was we achieve a backbone with significant stiffness against bending.

We account for the non-linear nature of the magnetisation of super-paramagnetic nanoparticles (NPs). In order to do this, accurate calculations of the total field $\vec{H}_{tot}$ in each point of the system are necessary. The total magnetic field is the sum of $\vec{H}$ and the dipole field $\vec{H}_{d}$ where the latter, created by particle $j$, at position $\vec{r}_0$, is given by:
\begin{equation}\label{eq:dip-field}
\vec{H}_{d}=\dfrac{3\vec{r}_{0j} \cdot \vec{\mu_j}}{r_{0j}^5}\vec{r}_{0j}-\dfrac{\vec{\mu_j}}{{r}_{0j}^3}.
\end{equation}
The dipole moment, $\vec{\mu}_i$, of an $i$-th super-paramagnetic particle with mass $m$ at a given temperature $T$, is defined as:
\begin{equation}\label{eq:dip-spm}
\vec{\mu_i}=\mu_{max} L\left(\frac{\mu_{max}|  \vec{H}_{tot}|}{k_B T}\right) \dfrac{\vec{H}_{tot}}{H_{tot}},
\end{equation}
where $\mu_{max}=|\vec{\mu}_{max}|$ denotes the modulus of the maximal magnetic moment of the particle, $\vec{\mu}_{max}$. $k_B$ is the Boltzmann constant and $L(\alpha)$ is the Langevin function:
\begin{equation}\label{eq:langevin}
L(\alpha)=\coth(\alpha)-\dfrac{1}{\alpha}.
\end{equation}
Expression \eqref{eq:dip-spm} is the generalisation of  mean-field approaches, such as the modified mean field approach,\cite{ivanov2001magnetic} with the important distinction that we do not make any assumptions when calculating $\vec{H}_{tot}$. This approach is verified by the analytical calculations for super-paramagnetic particle magnetisation.\cite{elfimova2019static} The two main interactions we are interested in, are the long-range magnetic inter-particle interactions and central attraction between the colloids. For the central attraction between the colloids, we use a Lennard-Jones potential as given in Eq.\eqref{eq:LJ} ($\sigma=1$), where we can adjust the strength of the interaction by taking different values of $\epsilon$. We account for the long-range magnetic inter-particle interactions by means of the conventional dipole-dipole pair potential:
\begin{equation}
U_{dd}(\vec r_{ij})=\frac{\vec{\mu}_{i}\cdot\vec{\mu}_{j}}{r^{3}}-\frac{3\left[\vec{\mu}_{i}\cdot\vec{r}_{ij}\right]\left[\vec{\mu}_{j}\cdot\vec{r}_{ij}\right]}{r^{5}},
\label{eq:dipdip}
\end{equation}
where $\vec \mu_i$ and $\vec \mu_j$ are their respective dipole moments, $\vec r_{ij} = \vec r_i - \vec r_j$ is the displacement vector connecting their centers and $r=\left | \vec r _{ij}\right |$. Because of the fact that we are interested in the competition between the dipole-dipole interaction and the central attraction, we have chosen to set the energy scale to be measured in units of thermal energy,  $k_b T=\epsilon=1$, which we equate with the energy scale of the inter-colloid repulsion interaction. Even though Van der Waals attraction originates from polarisation of the particle surfaces and has a dipolar nature, this interaction is very short-ranged and effectively central. It only affects the assembly of particles that are approaching close contact. In contrast, non-central magnetic dipole-dipole interaction, described by Eq. \eqref{eq:dipdip} is slowly decaying with growing inter-particle distance. As a result, the interplay of strengths and ranges of the two latter interactions makes the structural properties unique and worth investigating.

\begin{figure*}[!h]
	\centering
\subfigure[]{\label{fig:gyr_pln_1}\includegraphics[width=0.51\columnwidth]{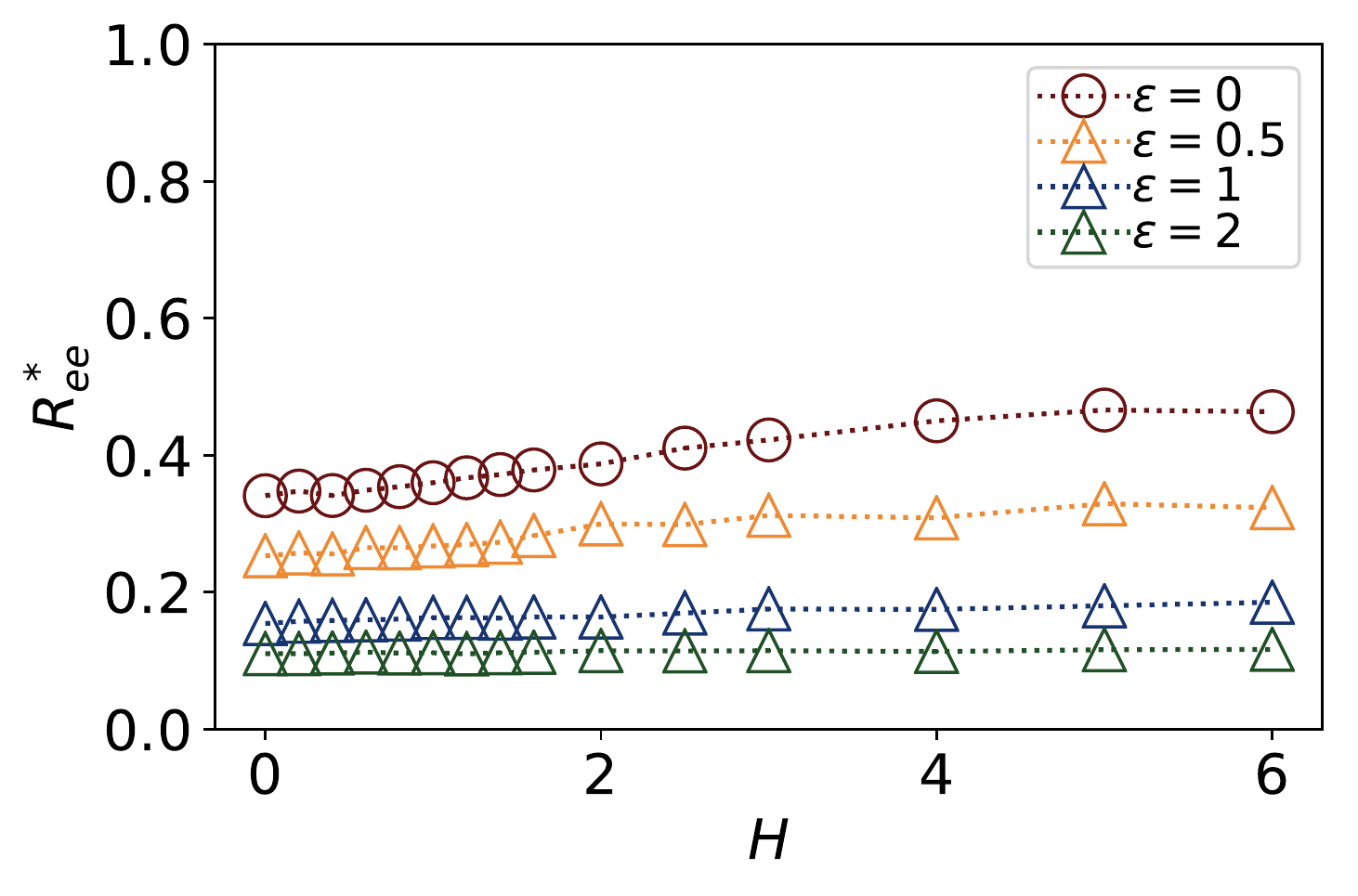}}
\subfigure[]{\label{fig:gyr_pln_3}\includegraphics[width=.51\columnwidth]{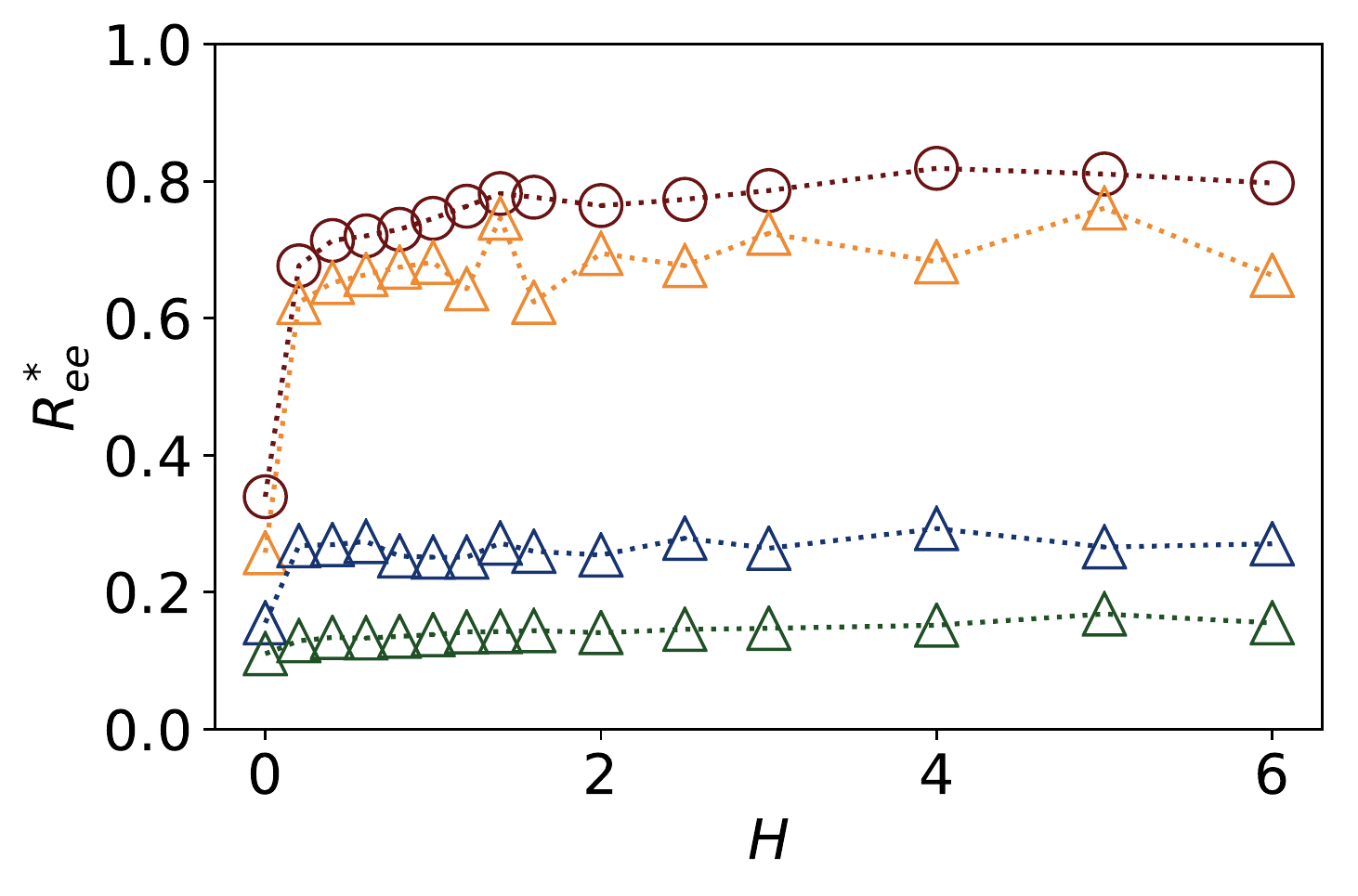}}
\subfigure[]{\label{fig:gyr_cnst_1}\includegraphics[width=.51\columnwidth]{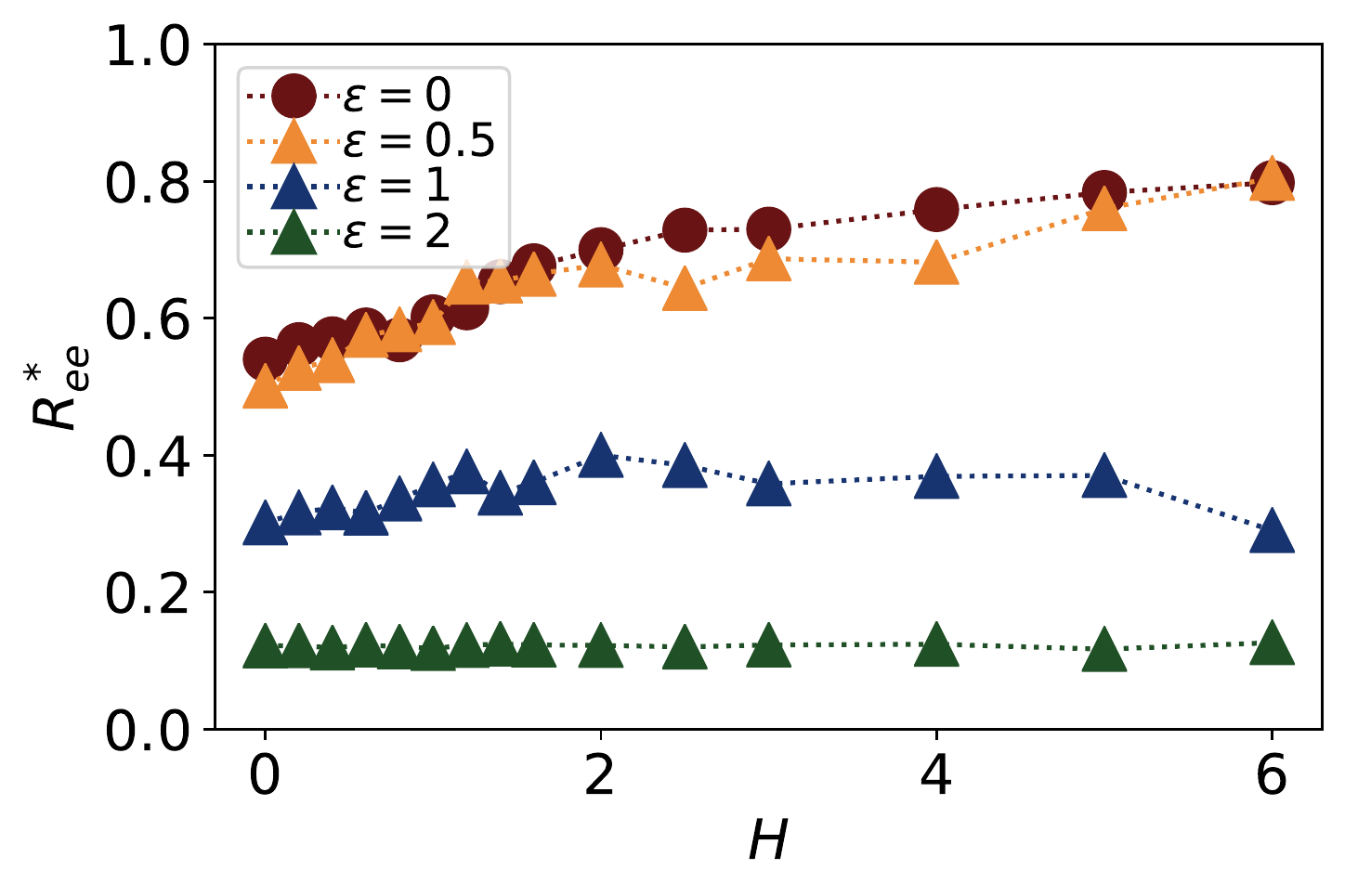}}
\subfigure[]{\label{fig:gyr_cnst_3}\includegraphics[width=.51\columnwidth]{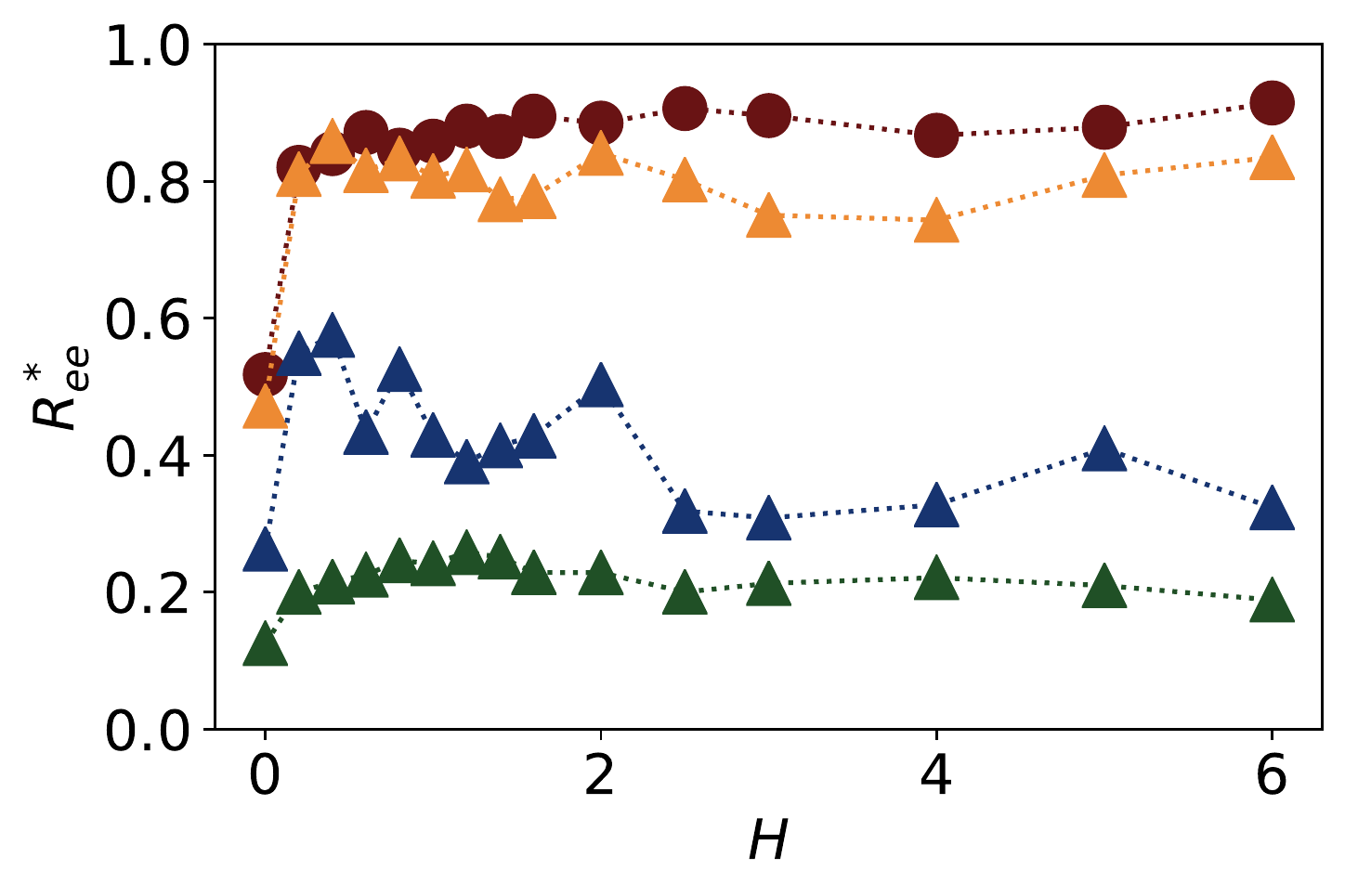}}
	\caption{End-to-end distance, $R_{ee}^*$, where $R_{ee}^*=R_{ee}/(N d)$, plotted against the applied external magnetic field $H$. Here $N$ is the number of monomers in a filaments, and $d$ is the equilibrium inter-particle distance.  Subplots (a) and (b) correspond to MFs with plain crosslinking;  Subplots (c) and (d) correspond to MFs with constrained crosslinking. In (a) and (c) $\mu_{max}^2=1$; in (b) and (d) $\mu_{max}^2=3$. Each subplot shows four curves, representing results for MFs super-paramagnetic NPs and various $\epsilon$ of the VdW interaction, as explained in the legend.}
	\label{fig:gyr}
\end{figure*}
 
We performed extensive MD simulations in the canonical ensemble for both of our crosslinking models, at different values dimensionless applied filed $H$, $\vec{\mu}_{max}$, and $\epsilon$. The simulations were done using ESPResSo. \cite{2006-limbach} Simulation of the background fluid was done using Langevin Dynamics (LD).\cite{allen2017computer} Long range dipole-dipole interactions were calculated using direct summation. We run a relaxation cycle for $10^{7}$ integration steps, after which we switch on $\vec{H}$ and start recording data. We use a time-step of $10^{-2}$ during the measurement run. We record instances during $1 050 000$ integration steps, where each recorded instance is separated by $3000$ integration's each. All results presented are made based on average of twenty independent initial configuration runs. Importantly, initial orientations of MFs in our simulations are uniformly distributed on a surface of a sphere. We use an iterative magnetisation procedure, described in Eq. \eqref{eq:dip-spm}. Computationally, after every integration of Eq. \eqref{eq:langevin}, we reevaluate $\vec{H}_{tot}$, and based the approach we described above, reassign new values of dipole moments to the magnetic colloids in the system.

\section{Results and Discussions}\label{sec-res}

We want to gain insight in to the equilibrium structure of super-paramagnetic filaments, in two distinct crosslinking approaches. As a first measure, we discuss how relative scale of the dipole-dipole interaction and central attraction affects normalised end-to-end distance, $R_{ee}^*$, of a MF subjected to an external magnetic field $H$. For further reference it should be noted that results which are plotted with hollow symbols correspond to filaments with plain crosslinking, while filled symbols correspond to constrained crosslinking.

As it can be seen in Fig. \ref{fig:gyr_pln_1}, for plain crosslinking and $\mu_{max}^2=|\vec{\mu}_{max}|^2=1$, $R_{ee}^*$ remains mostly flat as $H$ increases, regardless of the strength of central attraction given by $\epsilon$. Clearly in this case magnetic interactions are weak and are dominated by entropy. As we increase the depth of the central attraction potential, we reach even more compact states of MFs, as the system attempts to maximize the average number of neighbours for each monomer. For $\mu_{max}^2=3$, shown in Fig. \ref{fig:gyr_pln_3}, we can observe a sudden rise in $R_{ee}^*$ in the low field region ($H<1$) in otherwise mostly flat profiles, for $\epsilon=0$ and $\epsilon=0.5$. We maintain largely the same profiles as for $\mu_{max}^2=1$ for higher strengths of central attraction. Clearly, for $\mu_{max}^2=3$, the dipole-dipole interaction is strong enough to overcome entropy, even for $H<1$, and for $\epsilon \leq 0.5$, central attraction cannot compete. Therefore, we see MFs unravel in to a state approaching a head-to-tail dipole arrangement of neighboring monomers. Since MFs with plain crosslinking can also bend in an applied magnetic field without any loss of magnetic energy and significant gain of entropy, central attraction will further collapse such a bent conformation in order to maximise the "touching" of monomers. For $\epsilon \geq 1$, central attraction is strong enough to compete with the dipole-dipole interaction. Therefore, for plain crosslinking and $\mu_{max}^2=3$, we see hardly any change compared to what we have seen for plainly crosslinked MFs with $\mu_{max}^2=1$. The overall shape of these profiles we attribute to the local orientation of the dipole moments along the field direction. For high values of $H$, MFs with $\epsilon \leq 0.5$ reach a state when its $R_{ee}^*$ is approximately 20$\%$ lower than that of a straight rod. Otherwise, they remain below 30$\%$ of the end-to-end distance of a straight rod.

Once we look at $R_{ee}^*$ for MFs with constrained crosslinking, plotted in Figs.\ref{fig:gyr_cnst_1} and \ref{fig:gyr_cnst_3}, we note that, excluding the case where $\epsilon=2$, the degree of elongation is higher overall, for both values of $\mu_{max}^2$. For $\mu_{max}^2 = 1$, we attribute the increased elongation of MFs to the increased rigidity of the backbone. The effect is less pronounced however for $\mu_{max}^2 = 3$ because the dipole-dipole interaction is dominant. For $\epsilon=2$, central attraction is strong enough to win against the constrained crosslinking, leading to the field dependence of $R_{ee}^*$ similar to that found for MFs with plain crosslinking. For $\mu_{max}^2 = 1$, correlations introduced by the constrained crosslinking are enough to make the $R_{ee}^*$ for $\epsilon=0$ and $\epsilon=0.5$  indistinguishable for the whole range of $H$. In this case, an elongation of the MF  for $\epsilon=1$ can also be observed. 

Generally, the increased correlations between dipole orientations and the MF backbone leads to higher values of $R_{ee}^*$ in comparison to the plain model in Fig. ~\ref{fig:gyr_pln_3}. Energetically, as we will show below, this seems to be an opportune region for the central attraction to force strange "compromise" configurations on to the dipoles. 

\begin{figure}[h!]
	\centering
\subfigure[]{\label{fig:neighbours_pln_mu1_vdw}\includegraphics[width=0.49\columnwidth]{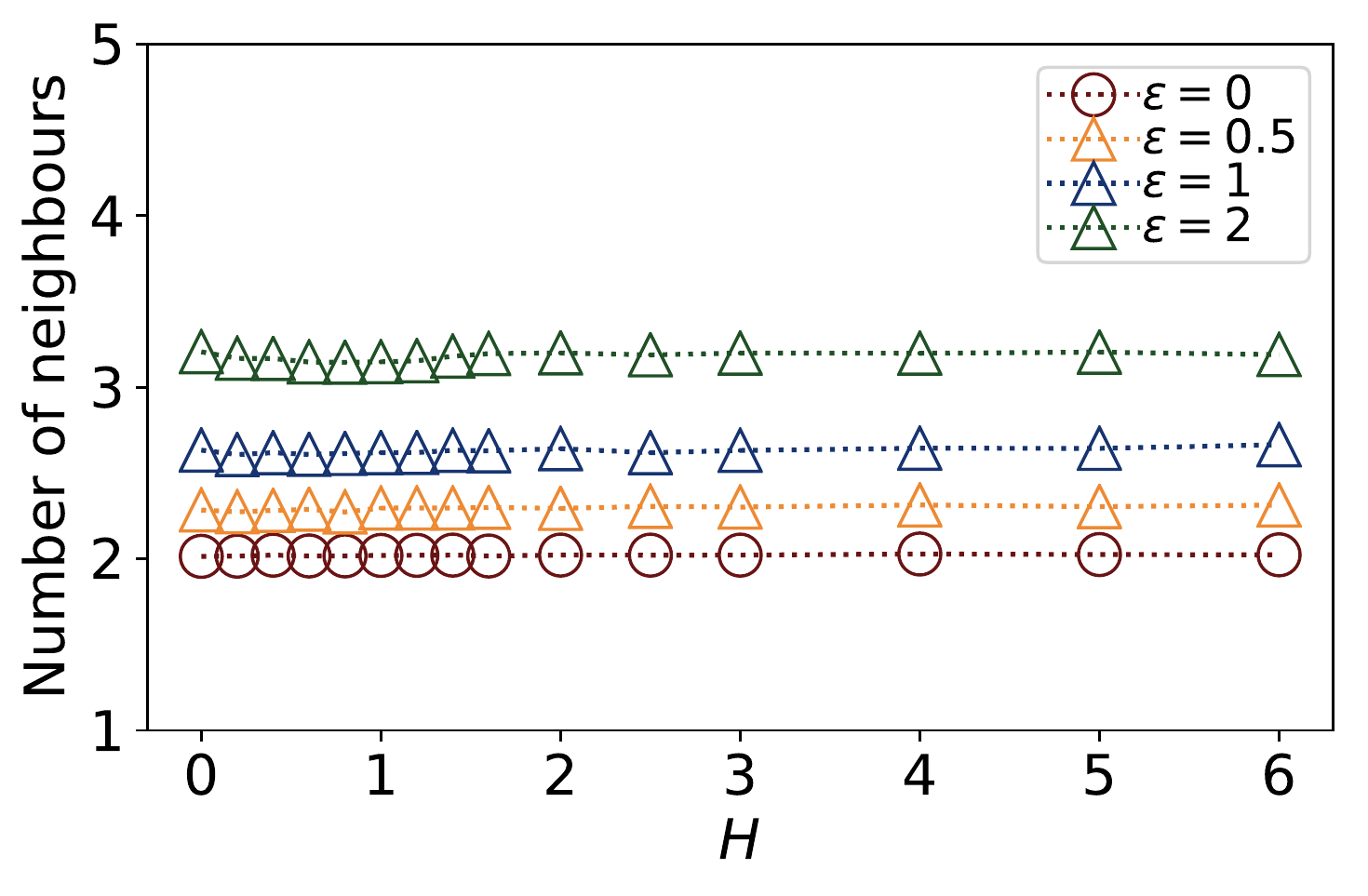}}
\subfigure[]{\label{fig:neighbours_pln_mu3_vdw}\includegraphics[width=0.49\columnwidth]{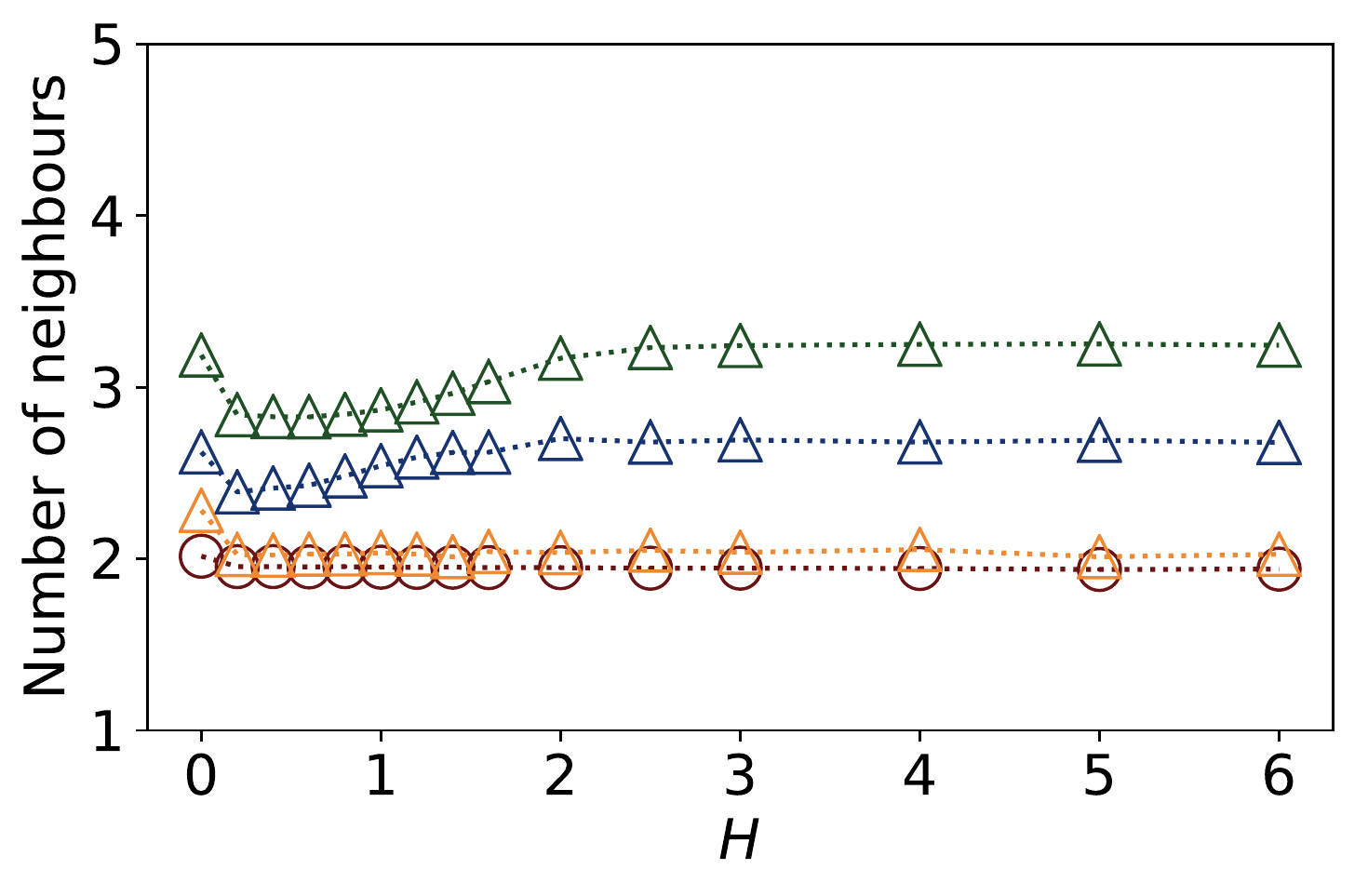}}
\subfigure[]{\label{fig:neighbours_cnst_mu1_vdw}\includegraphics[width=0.49\columnwidth]{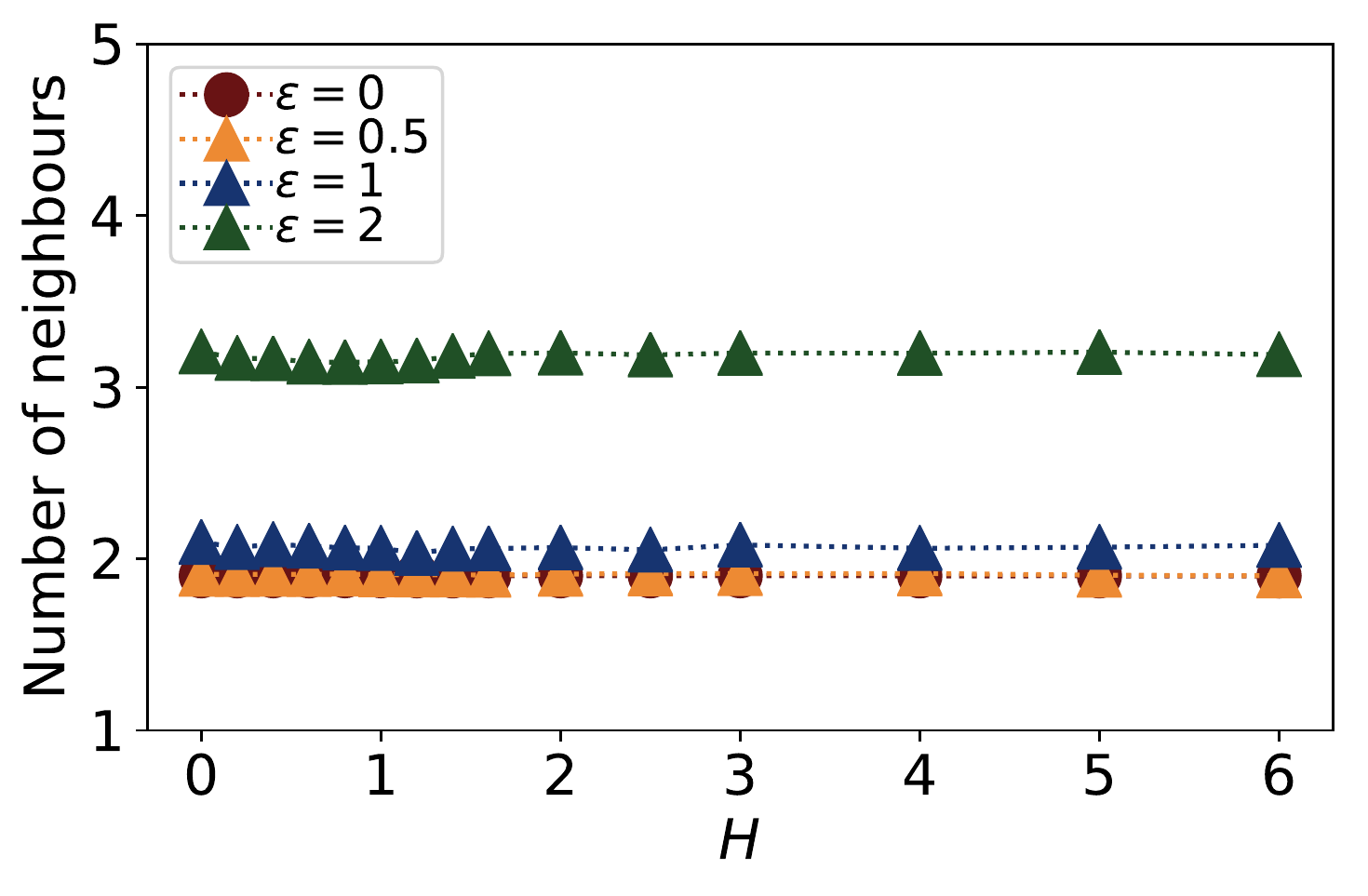}}
\subfigure[]{\label{fig:neighbours_cnst_mu3_vdw}\includegraphics[width=0.49\columnwidth]{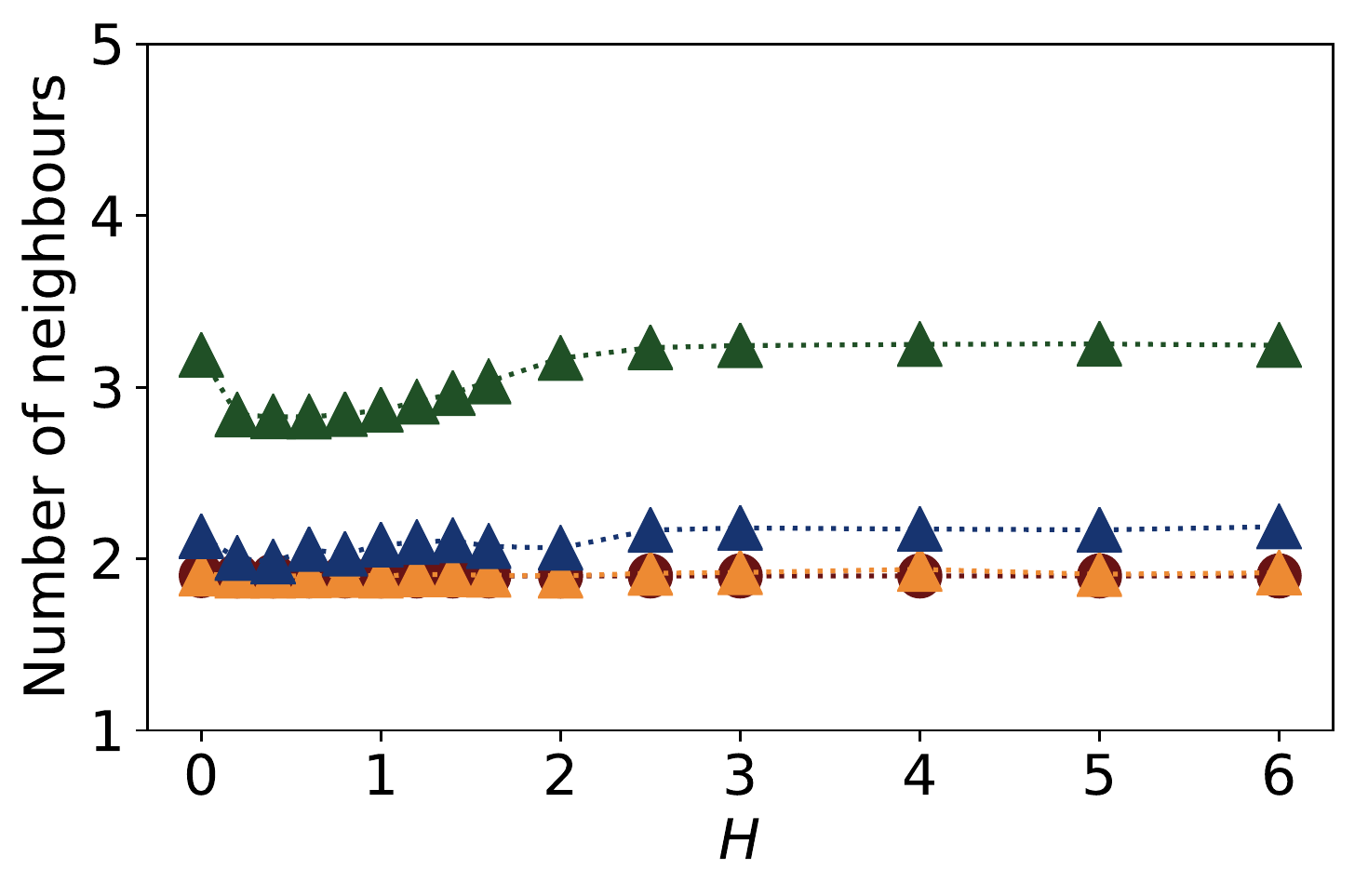}}
	\caption{Number of neighbours {\it versus} magnetic field strength $H$.  Subplots (a) and (b) correspond to MFs with plain crosslinking; Subplots (c) and (d) correspond to MFs with constrained crosslinking. In (a) and (c) $\mu_{max}^2=1$; in (b) and (d) $\mu_{max}^2=3$. Each subplot shows four curves, representing results for MFs super-paramagnetic NPs and various $\epsilon$ of the VdW interaction.}
	\label{fig:neighbours}
\end{figure}

So, to summarise, in terms $R_{ee}^*$ curves, constrained crosslinking not only increases the overall values of $R_{ee}^*$, but also unveils a complex interplay between magnetic and central attraction forces, which seem to be highlighted for less rigid MFs with $\mu_{max}^2 = 3$ and $\epsilon=1$. None of the MFs studied here stretches to its full length under the influence of $H$. 

\begin{figure}[!h]
	\centering
\subfigure[]{\label{fig:shape_anisotropy_pln_mu1_vdw}\includegraphics[width=0.49\columnwidth]{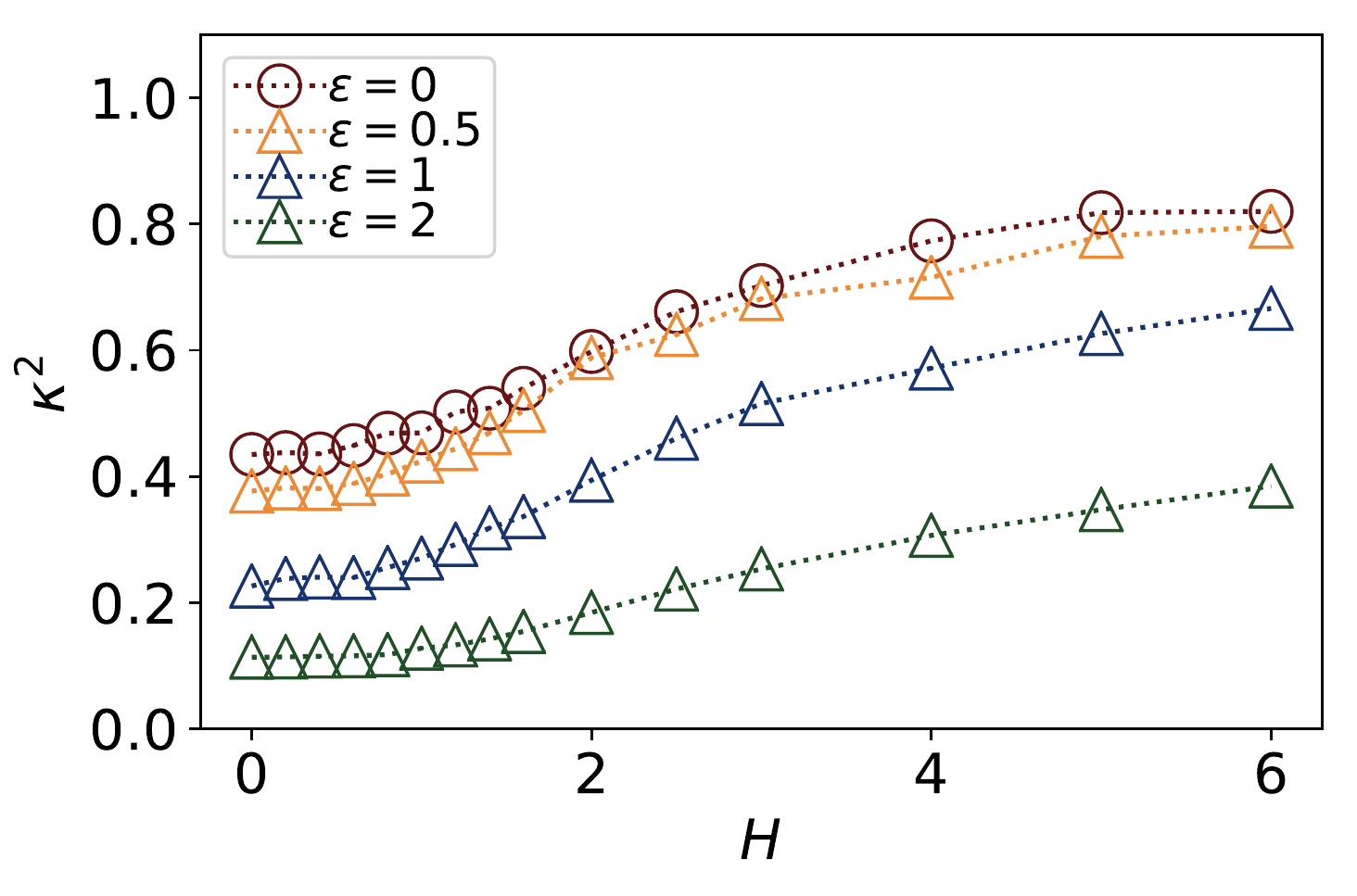}}
\subfigure[]{\label{fig:shape_anisotropy_pln_mu3_vdw}\includegraphics[width=.49\columnwidth]{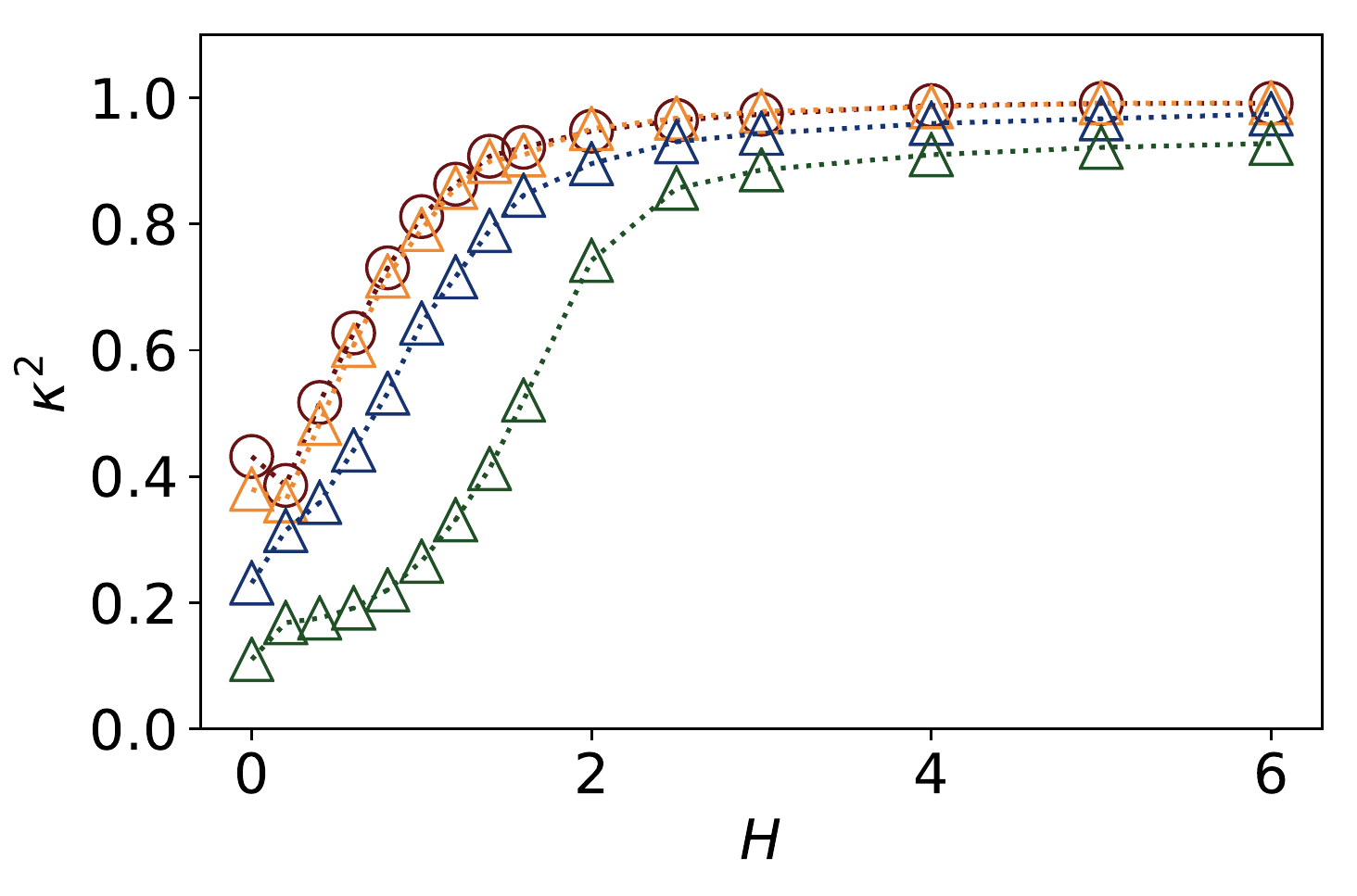}}
\subfigure[]{\label{fig:shape_anisotropy_cnst_mu1_vdw}\includegraphics[width=.49\columnwidth]{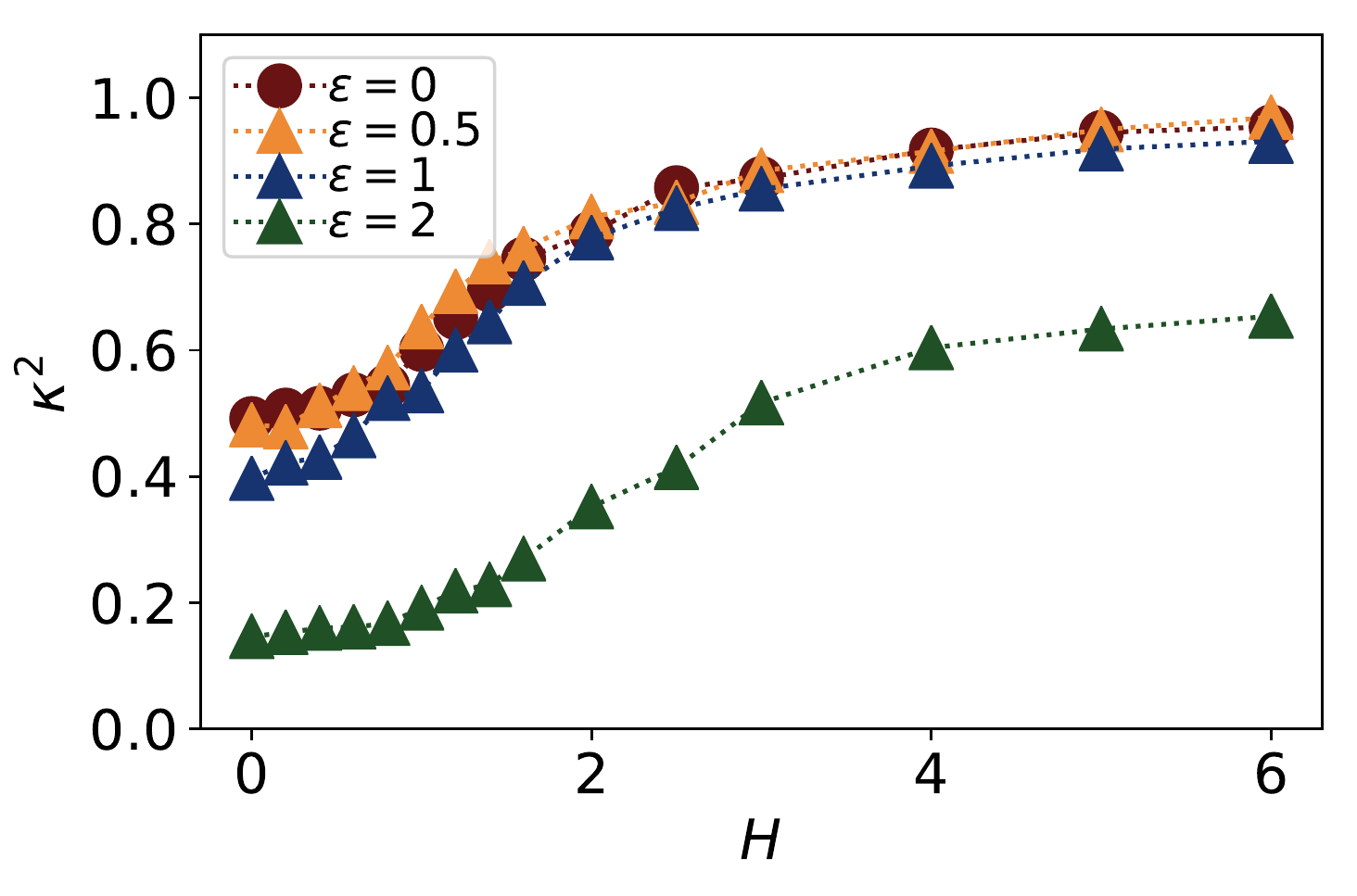}}
\subfigure[]{\label{fig:shape_anisotropy_cnst_mu3_vdw}\includegraphics[width=.49\columnwidth]{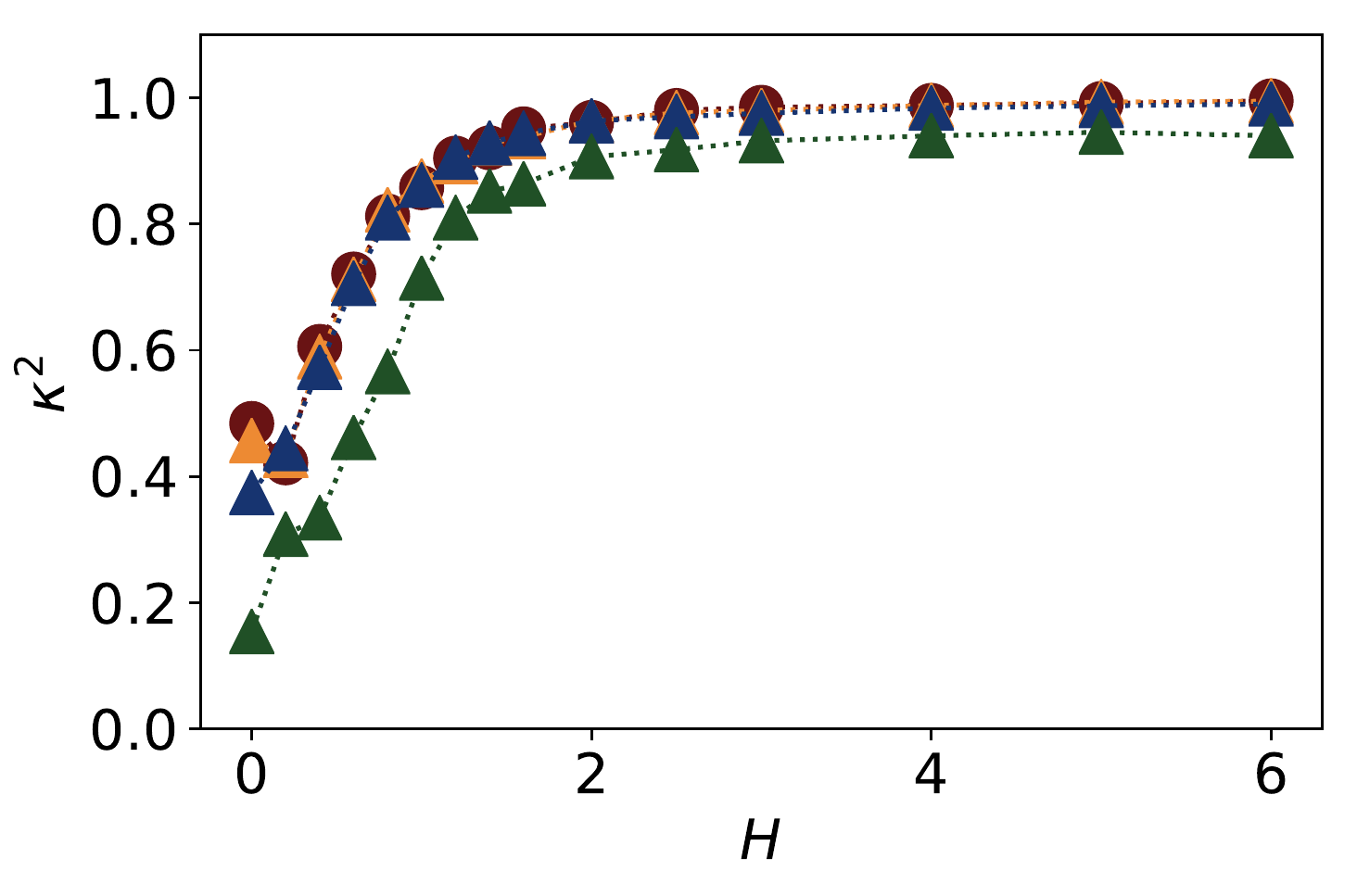}}
	\caption{Relative shape anisotropy, $\kappa^2$, Eq. \eqref{eq:kappa}, plotted against the applied external magnetic field $H$. Subplots (a) and (b) correspond to MFs with plain crosslinking; Subplots (c) and (d) correspond to MFs with constrained crosslinking. In (a) and (c) $\mu_{max}^2=1$; in (b) and (d) $\mu_{max}^2=3$. Each subplot shows four curves, representing results for MFs super-paramagnetic NPs and various $\epsilon$ of the VdW interaction, as explained in the legend.}
	\label{fig:shape_anisotropy}
\end{figure}

To further elucidate the structural properties of super-paramagnetic MFs with a focus on the joined effect of central attraction, dipole interactions and crosslinking, we consider how does the average number of neighbours, each colloid has, vary as a function of $H$. This dependence is shown in Fig.  \ref{fig:neighbours}. Furthermore, in conjunction with the average number of colloidal neighbours, we can consider the relative shape anisotropy of a MF, $\kappa^2$, as a function of $H$, as shown in Fig. \ref{fig:shape_anisotropy}. The shape anisotropy $\kappa^2$ is defined as 

\begin{equation}\label{eq:kappa}
\kappa^2=\dfrac{1}{2}\dfrac{\lambda_x^4+\lambda_y^4+\lambda_z^4}{(\lambda_x^2+\lambda_y^2+\lambda_z^2)^2}-\dfrac{3}{2},
\end{equation}
\noindent where $\lambda_x,\lambda_y,\lambda_z$ are the principal moments of the gyration tensor. The gyration tensor, ${\mathbf R_{ab}}$, is defined as
\begin{equation}
{\mathbf R_{ab}}=\dfrac{1}{N}\sum_{i=1}^Nr_a^{(i)}r_b^{(i)},
\end{equation}
\noindent where $r_a^{(i)}$ is the $a$-th Cartesian coordinate of of $i$-th particle in in the center-of mass reference frame. This parameter provides a more complete understanding of the $R_{ee}^*$ tendencies, we have seen in Fig. \ref{fig:gyr}.

Looking at average number of colloidal neighbours for $\mu_{max}^2 = 1$ in Figs. \ref{fig:neighbours_pln_mu1_vdw} and \ref{fig:neighbours_cnst_mu1_vdw}, we see that the strength of external magnetic field $H$ has virtually no significance, regardless of crosslinking. Furthermore, we can stipulate that, with the noted exception of curves for $\epsilon=0.5$ and $\epsilon=1$, the profiles are rather similar for both crosslinking approaches. MFs without central attraction remain in a non-collapsed, albeit rather coiled state, across the applied magnetic field range. On the other hand, for $\epsilon=2$, MF conformations collapse and should resemble structures with spherical symmetry. For $\epsilon=0.5$ and $\epsilon=1$ however, we see a pronounced drop in the average number of colloid neighbours for constrained crosslinking, which together with the increase in $R_{ee}^*$, see Figs. \ref{fig:gyr_pln_1} and \ref{fig:gyr_cnst_1}, suggest that the structures MFs with central attraction of strength $\epsilon=0.5$ and $\epsilon=1$ exhibit more elongated, cylindrically symmetric structures for constrained crosslinking, as compared to plain crosslinking. Based on the number of neighbours is seems that the increased rigidity of the backbone, due to constrained crosslinking, prevents collapsed conformations of MFs, which would be the preferred ones to maximise the impact of central attraction.
For filaments with $\mu_{max}^2 = 3$, shown in Figs.\ref{fig:neighbours_pln_mu3_vdw} and \ref{fig:neighbours_cnst_mu3_vdw}, we also see a drop in the average number of neighbours, once $H$ is applied, because for $\mu_{max}^2 = 3$, Zeeman energy is rather significant already for $H\leq 1$.  The behaviour of the relative shape anisotropy $\kappa^2$, shown in Fig. \ref{fig:shape_anisotropy}, confirms what we have suggested from analysing $R_{ee}^*$ and average number of nearest neighbours. Increased inter-particle correlation due to constrained crosslinking and/or increase in $\mu_{max}^2$ lead to an increase in $\kappa^2$, which can be explained by elongation of the conformations of MFs in field direction. In other words, initially spherical conformations of MFs, due to the isotropic central attraction, change to cylindrically symmetric structures, due to increased Zeeman coupling and as a consequence, dipole-dipole forces.

\begin{figure*}[!ht]
	\centering
\subfigure[]{\label{fig:magnet_plain_lambda1}\includegraphics[width=0.75\columnwidth]{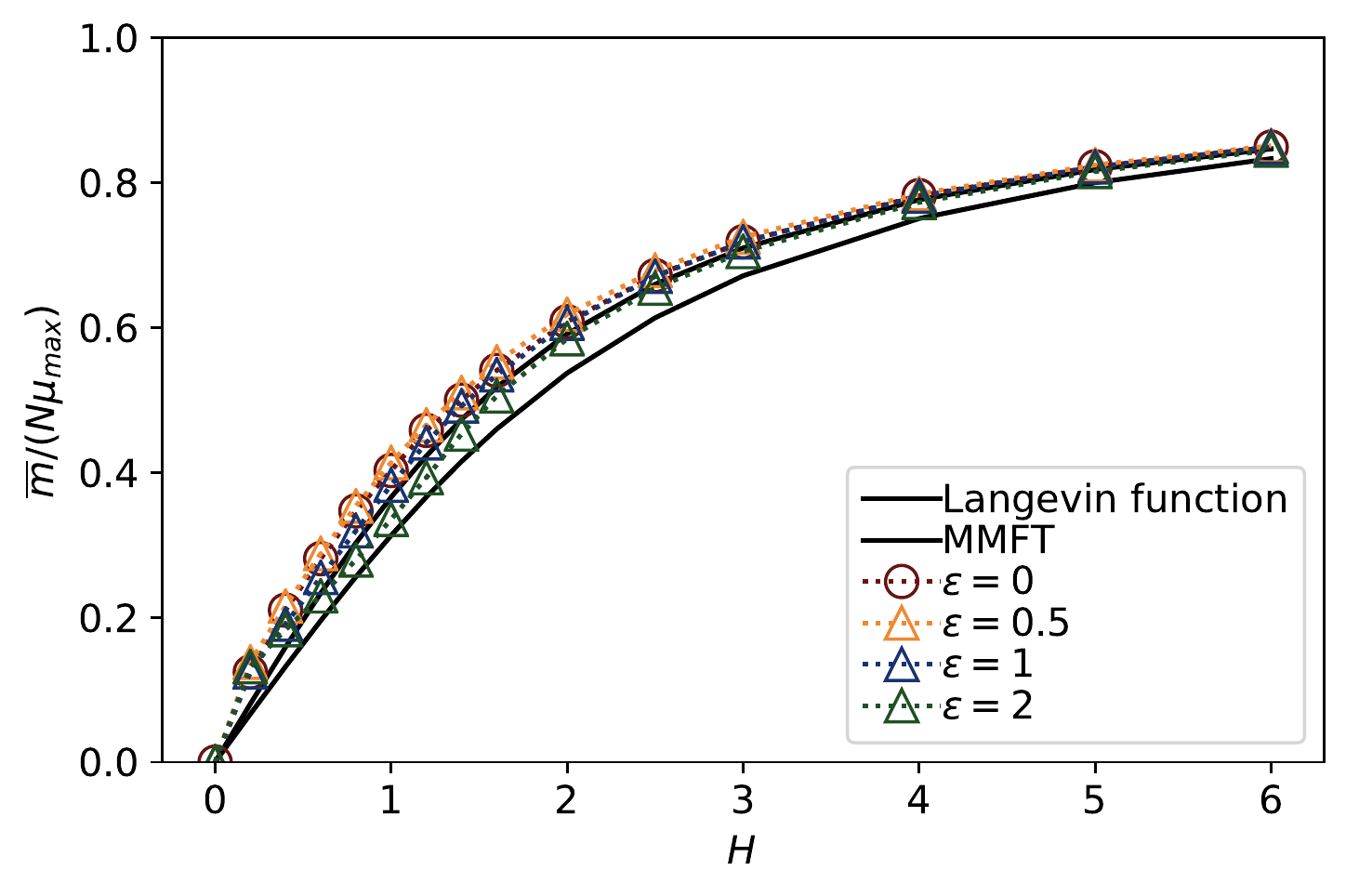}}
\subfigure[]{\label{fig:magnet_plain_lambda3}\includegraphics[width=.75\columnwidth]{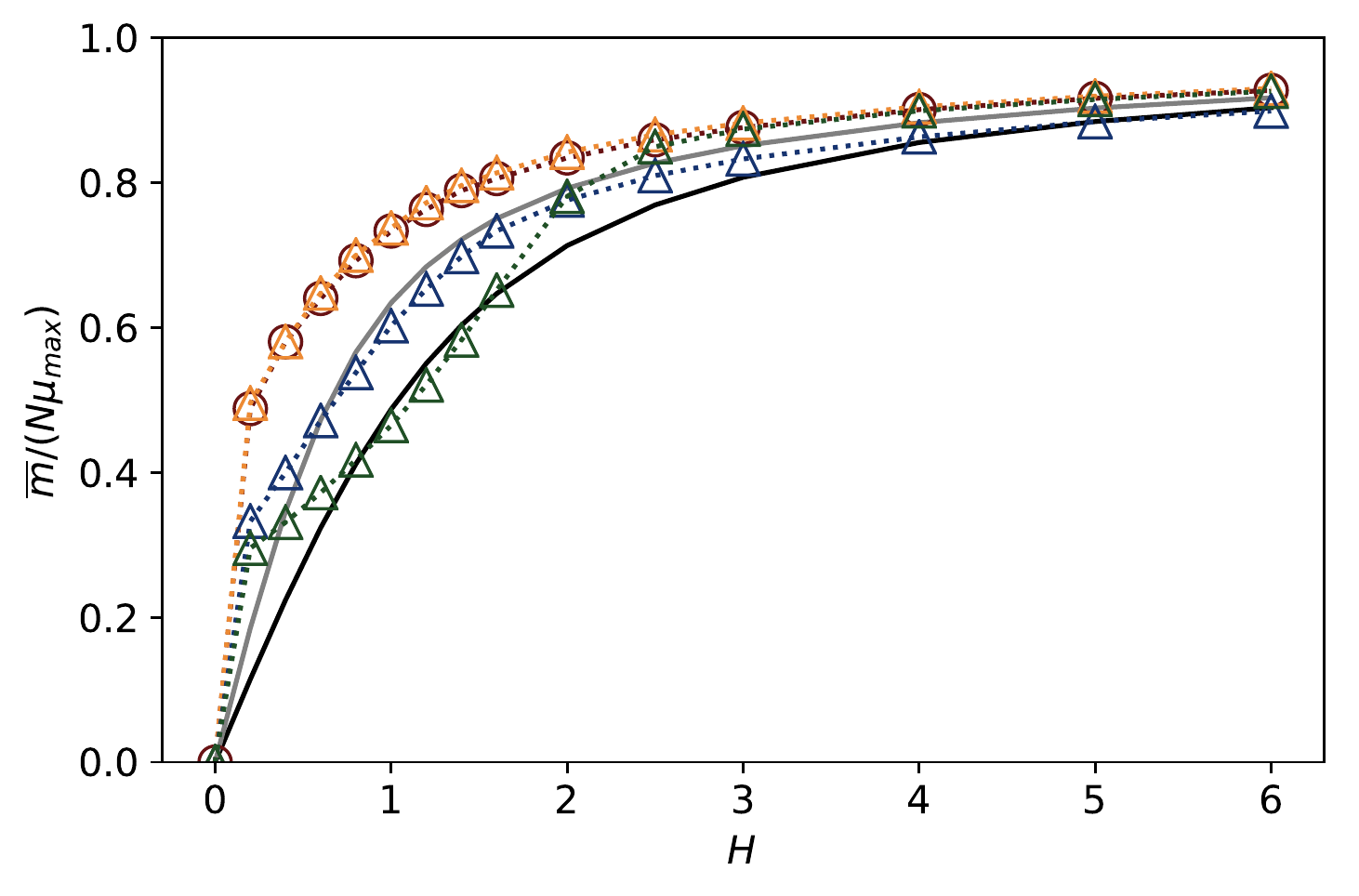}}
\subfigure[]{\label{fig:magnet_cnst_lambda1}\includegraphics[width=.75\columnwidth]{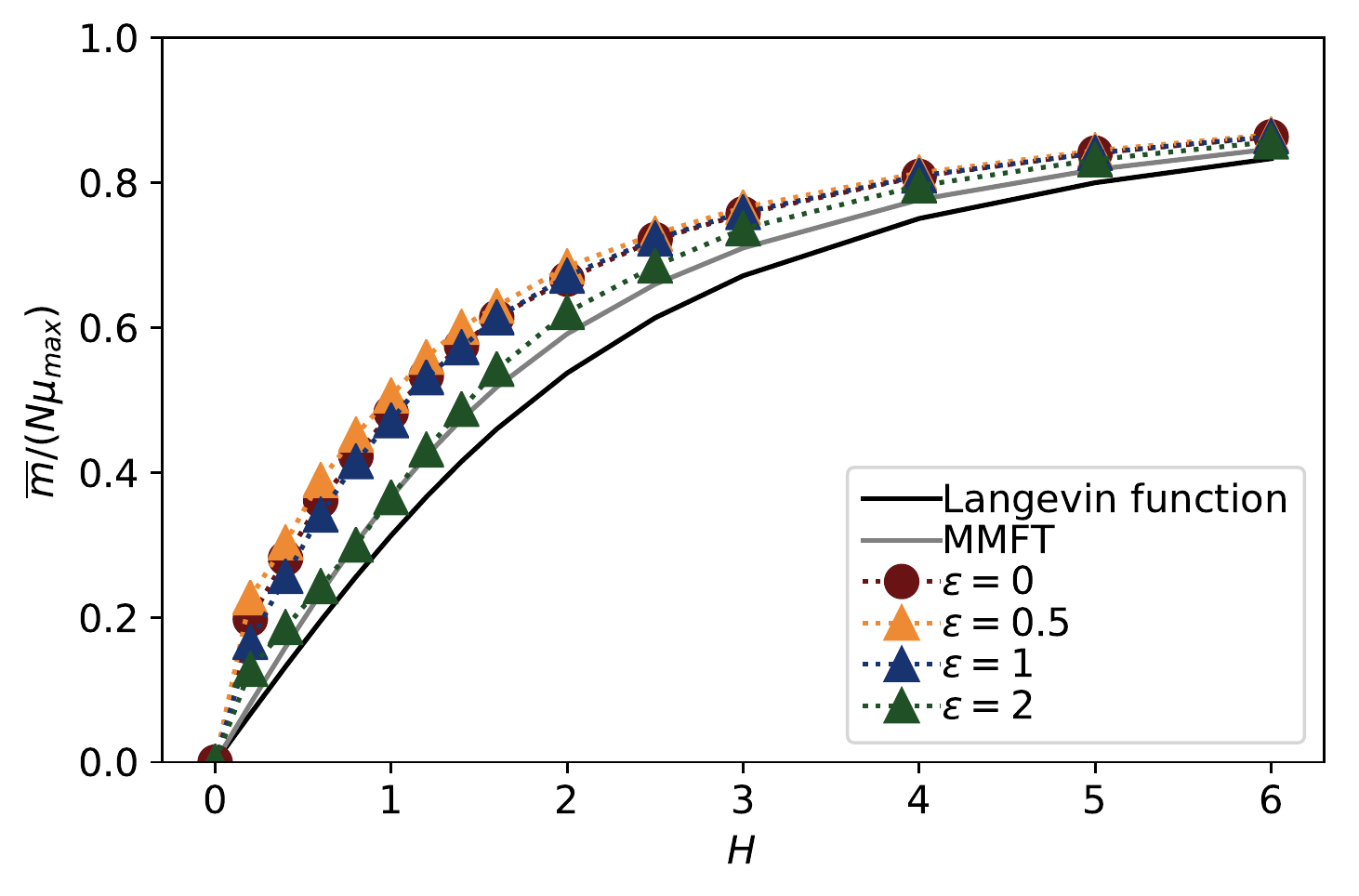}}
\subfigure[]{\label{fig:magnet_cnst_lambda3}\includegraphics[width=.75\columnwidth]{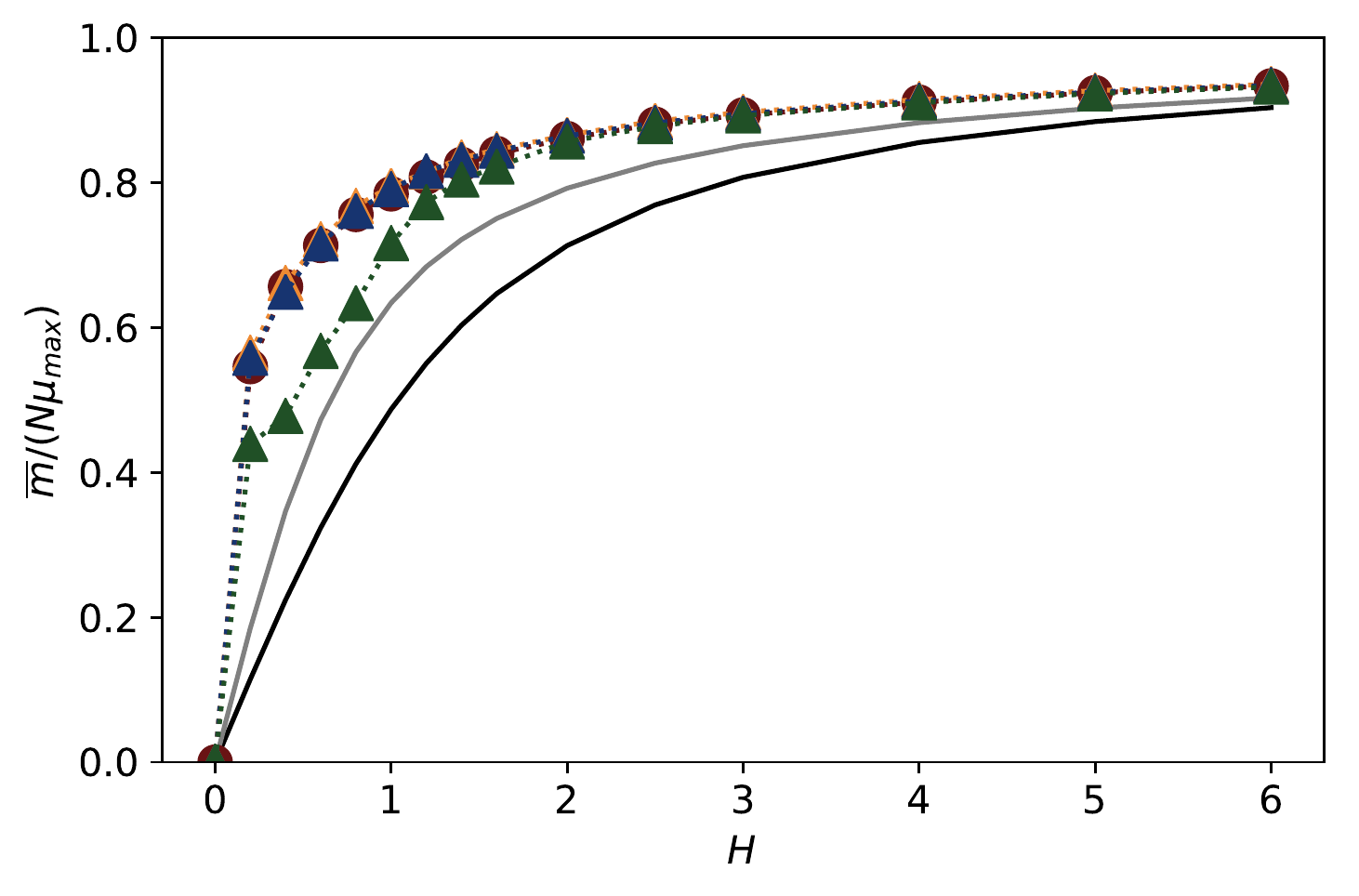}}
	\caption{Projection of the total filament magnetic moment, normalised by $N\mu_{max}$, on $\vec{H}$ {\it versus}  magnetic field strength $H$. Subplots (a) and (b) correspond to MFs with plain crosslinking; Subplots (c) and (d) correspond to MFs with constrained crosslinking. In (a) and (c) $\mu_{max}^2=1$; in (b) and (d) $\mu_{max}^2=3$. Each subplot shows four curves, representing results for MFs with super-paramagnetic NPs and various $\epsilon$ of the VdW interaction, Langevin magnetisation law from Eq. \eqref{eq:langevin} and MMFT from Eq. \eqref{eq:mmft} (see the legend).}
	\label{fig:pros10}
\end{figure*}

As a measure of the overall magnetic response of super-paramagnetic MFs with and without central attraction, we use the average of the normalised projection of the magnetic moment of a filament $\bar{m}$, on the direction of $H$, shown in Fig. ~\ref{fig:pros10}. The value $\bar{m}$ is normalised by  $N \mu_{max}$. For MFs where the monomers are non-interaction, the field dependence of normalised $\bar{m}$ should follow Eq.\eqref{eq:langevin}, as a function of $\alpha =\mu_{max}H$, previously introduced as the Langevin function. In each subfigure of Fig. \ref{fig:pros10}, $\bar{m}$, as estimated by Eq.\eqref{eq:langevin}, is shown with a black, solid line.
As we saw in Fig. \ref{fig:gyr}, $R_{ee}^*$ for $\mu_{max}^2 = 1$ of MFs is basically independent from $H$, regardless of $\epsilon$. Therefore, in this case the magnetic response of a MF should be equivalent to a response of the same number of non-crosslinked particles, at a given particle density. We substantiate this by plotting $m_T$ calculated via modified mean-field theory of the second order (MMFT) \cite{ivanov2001magnetic}, Eq. \eqref{eq:mmft}. MMFT should describe static magnetic properties of relatively concentrated ferrofluids well, assuming that dipole forces do not lead to formation of clusters.\cite{2007-ivanov} In the framework of MMFT, the magnetisation of a monodisperse system has the form: 
\begin{equation}\label{eq:mmft}
m_T=\rho^\ast\mu_{max} L\left(\mu_{max} H_e\right)
\end{equation}
$$
    H_e = H+\dfrac{1}{3}\mu_{max}\rho^\ast L(\mu_{max}H)+$$
$$+\dfrac{1}{48}\left(\mu_{max}\rho^\ast\right)^2 L(\mu_{max}H)\dfrac{dL(\mu_{max}H)}{dH}.
$$

\noindent Here,  $\rho^*$ is the number density of MNPs. We assume that, for a given value of $H$, all $N=20$ MNPs are located within a volume $V = 4\pi R_g^3/3$, so that $\rho = N/V$. Reduced magnetisation, $m_T$, calculated via MMFT is given by a solid gray line in each subfigure of Fig. \ref{fig:pros10}. 

For plainly crosslinked MFs and $\mu_{max}^2 = 1$, as can be seen on Fig.  \ref{fig:magnet_plain_lambda1}, the contribution of the dipole field to the magnetisation of super-paramagnetic NPs is significant enough that the actual $\bar{m}$ is underestimated by the Langevin prediction. The results of Eq. \eqref{eq:mmft}, however, do describe $\bar{m}$ well. This means that under these conditions, the crosslinking does not affect the magnetic response of the individual particles. Across the range of $\epsilon$ we have, for plain crosslinking and $\mu_{max}^2 = 1$, we see very little influence of the central attraction on $\bar{m}$. Curves for $\epsilon=0$ and $\epsilon=0.5$ are indistinguishable, while we can notice only a slight drop in $\bar{m}$, as we increase the strength of central attraction to $\epsilon=1$ and $\epsilon=2$. Once we move to constrained crosslinking for $\mu_{max}^2=1$, shown in Fig. \ref{fig:magnet_cnst_lambda1}, the effects of central attraction become more pronounced. Firstly, we can note that $\bar{m}$ is well above MMFT predictions. Furthermore constrained crosslinking clearly enhances $\bar{m}$, especially in the $H\leq 1$ region. Inter-particle correlations introduced by the crosslinking enhance the magnetic response of super-paramagnetic MFs, where for values of $\epsilon\leq 1$, we cannot distinguish one from another. For $\epsilon=2$ however, we see the central attraction make $\bar{m}$ very close to what would be predicted by MMFT, as it tries to make chain conformations collapse, minimizing the steric part of the energy, but hindering the orientations of the dipole moments. For $\mu_{max}^2 =3$, we see a rather predictable increase in $\bar{m}$ for constrained crosslinking and $\epsilon\leq 1$, shown in Fig. \ref{fig:magnet_cnst_lambda3}. For stronger central attraction ($\epsilon=2$), we can clearly see the interplay between the dipole-dipole interaction and central forces, as $\bar{m}$ has a distinct dip in the $0.1\leq H\leq 1$ region. 
For plain crosslinking and $\mu_{max}^2 = 3$, magnetisation curves for $\epsilon=0$ and $\epsilon=0.5$ behave similarly to their counterparts for MFs with constrained crosslinking, albeit the overall $\bar{m}$ is slightly lower. The dip in magnetisation that we found for constrained crosslinking is also present for plain crosslinking, however it is much more pronounced. Central attraction clearly dominates the behaviour due to the inter-particle correlations introduced by plain crosslinking. 
\begin{figure}[!ht]
	\centering
\subfigure[]{\label{fig:snap-le}\includegraphics[width=0.26\columnwidth]{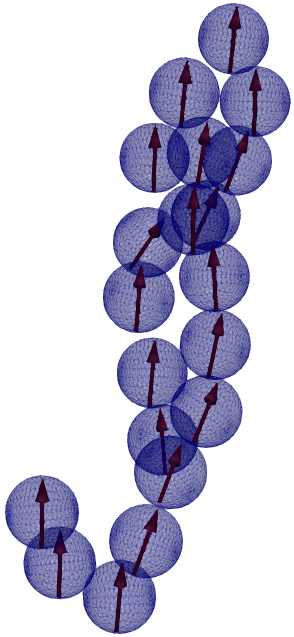}}
\subfigure[]{\label{fig:snap-he}\includegraphics[width=.24\columnwidth]{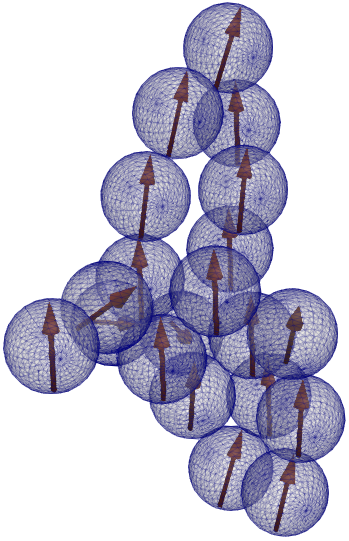}}
	\caption{Snapshots. (a) Low energy, less probable configuration; (b) High energy, more probable configuration. $\mu_{max}^2 = 3$, $\epsilon =1$, $H=1$.}
	\label{fig:snaps}
\end{figure}
Nonetheless, as Zeeman energy increases with growing $H$, dipole-dipole interaction prevails, which is consistent with what we have seen before. However, $\bar{m}$ curve for $\epsilon=1$, is found to be significantly below the curves for all other $\epsilon$, even in the high field region. Surprisingly even MMFT predicts a higher $\bar{m}$. In the region where $H \leq 1$, it is expected that the central attraction will compete with the dipole-dipole interaction, and, as we have seen before, lead to a lower net $\bar{m}$. But this being the case for $H \geq 4$ is rather unexpected. Given that we use averages in 20 independent initial configurations, we decided to inspect how do the $\bar{m}$ curves, as shown on Fig. \ref{fig:magnet_plain_lambda3} look for each of the separate initial configuration runs. Upon inspection, we recognised that there are in fact two distinct $\bar{m}$ profiles occurring in the case of plain crosslinking for $\mu_{max}^2 = 3$ and $\epsilon=1$: one is following the curve for $\epsilon=0.5$ and is met in approximately 20$\%$ of realisations, whereas the other curve has a very pronounced dip and is realised in the rest 80$\%$ of the realisations.

\begin{figure}[!ht]
	\centering
\subfigure[]{\label{fig:hist-mu1}\includegraphics[width=0.8\columnwidth]{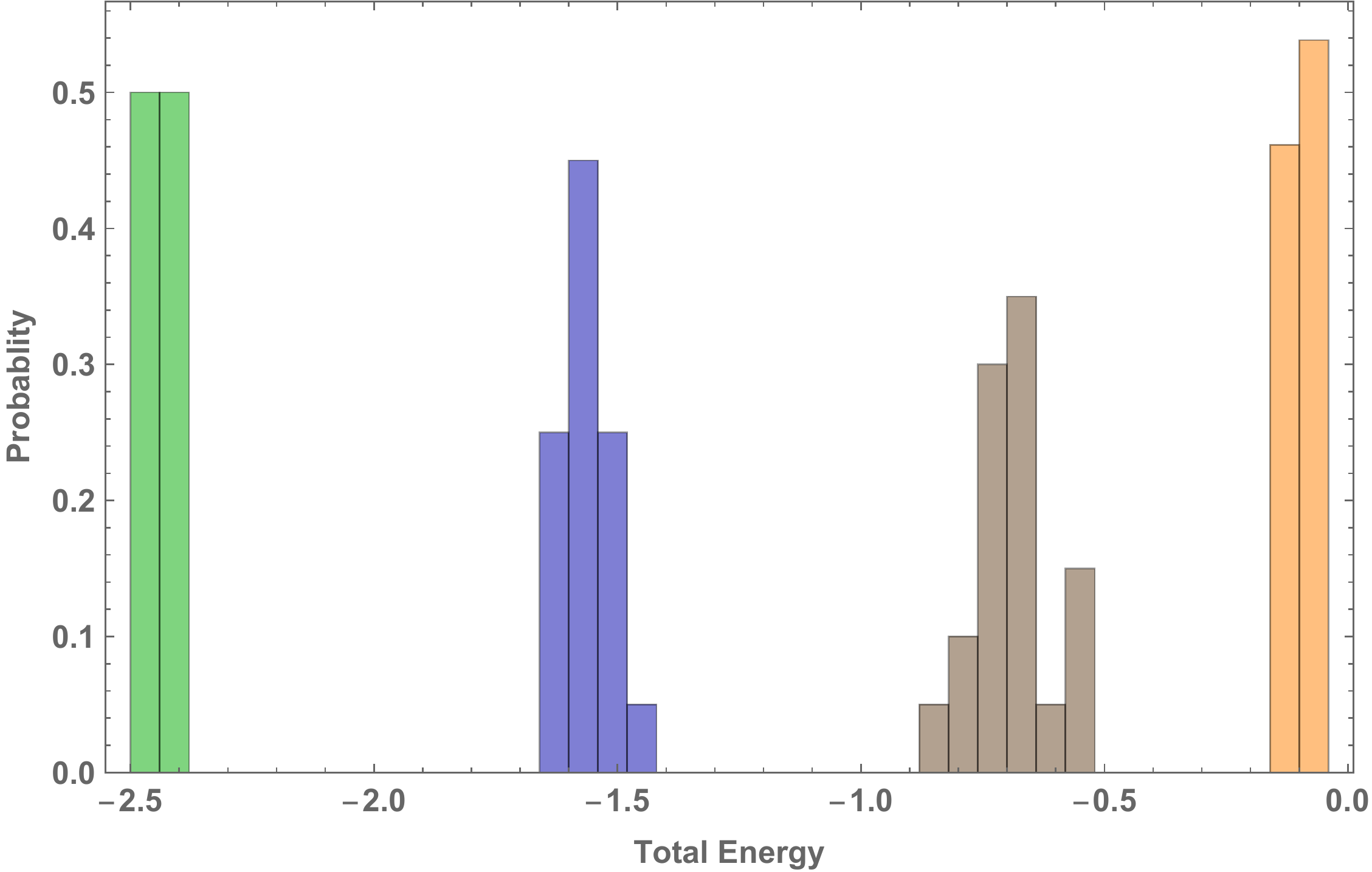}}
\subfigure[]{\label{fig:hist-mu3}\includegraphics[width=0.8\columnwidth]{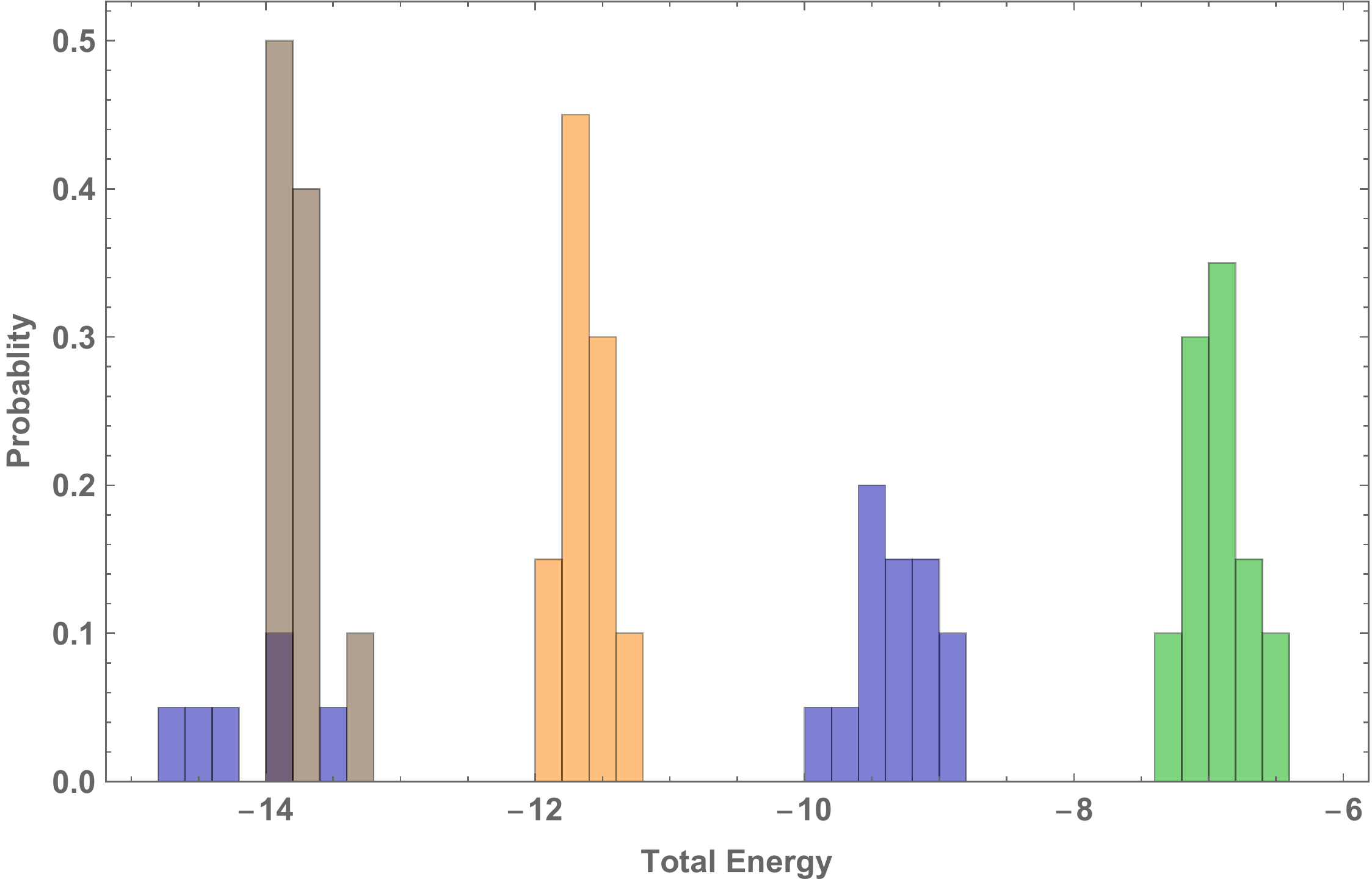}}
	\caption{Histograms, showing the probability of a filament to have a given total energy per particle. Both subfigures depict results for plain crosslinking, at $H=1$. Subfigure (a) corresponds to $\mu_{max}^2 = 1$; Subfigure (b) corresponds to $\mu_{max}^2 = 3$. Colors correspond to different values of $\epsilon$: brown -- $\epsilon=0$, orange -- $\epsilon=0.5$, blue -- $\epsilon=1$, green -- $\epsilon=2$. Only bimodal distribution is observed for $\mu_{max}^2 = 3$, $\epsilon =1$.}
	\label{fig:hist}
\end{figure}
As long as we observed two types of magnetisation, we decided to take a closer look at filament conformations corresponding to the two regimes and present them in Fig.  \ref{fig:snaps}. Here, the less frequent configuration is shown in Fig.  \ref{fig:snap-le} and the more frequent one -- in Fig.  \ref{fig:snap-he}.

In order to explain the origins of the two conformations, we calculated the total energy per particle for plain crosslinking and plotted this on two histograms in Fig. \ref{fig:hist}. Here, the total energy is the sum of Van der Waals, dipolar, Zeeman and elastic terms. As it can be seen in Fig. \ref{fig:hist-mu3}, all but one configurations have a rather pronounced peak, corresponding to a well-defined energy minimum. The only bimodal energy distribution is found for plainly crosslinked filaments with $\mu_{max}^2 = 3$, $\epsilon =1$. For the configurations which result in lower total energy, and occur less frequently, shown in Fig. ~\ref{fig:snap-le}, the dipolar energy is minimised, while the LJ-contribution is not optimised sufficiently. The configurations which lead to a higher total energy, and which occur more frequently, shown in Fig. ~\ref{fig:snap-he}, have the LJ-energy term minimised, while the dipolar term appears to be higher. All the other energies remain comparable. The difference in probability of occurrence can be attributed to the higher entropy of the states characterised by lower LJ-energy. 

Clearly, much work remains to be done to understand this finding, but we are tempted to speculate, that with these parameters, we might be in the vicinity of a critical point for a vapor-liquid transition, inherent to Stockmayer fluids. Stockmayer particles experience a short-range, isotropic attraction in addition to dipolar forces. It is a well known fact that Stockmayer fluids undergo a vapour-liquid phase transition on cooling and concentration increase.\cite{moore2015liquid,van1993makes,panagiotopoulos1992direct,stevens1995structure,adams1981static,novak2019structure} Even though we are dealing with super-paramagnetic particles, and the direct comparison of the parameter space is not possible, indications of a critical point are present. It is worth mentioning, that the suspension of Stockmayer MFs with ferromagnetic particles does phase separate.\cite{novak2019structure, 2016-cerda-pccp}

\section{Conclusions} \label{sec-conc}

In this manuscript we employed MD simulations to investigate the interplay between competing magnetic and steric interactions on the conformations of a single magnetic filament with super-paramagnetic monomers. We studied two different crosslinking mechanisms: in one case, referred to throughout this manuscript as plain crosslinking, the filament in zero applied magnetic field corresponded to a freely-joint polymer chain; in the other, so-called constrained crosslinking case, additionally to the connections between monomer centres, we introduced a bending penalty, thus modelling the backbone rigidity. In our simulations, super-paramagnetism of magnetic colloids was accounted for in a fully accurate, nonlinear manner. We studied MFs with two distinct values of colloid magnetic saturation: dipolar interactions between weakly magnetic monomers in the saturation field are comparable to thermal fluctuations; for monomers with high saturation magnetisation, dipolar strength could exceed thermal fluctuations by the factor of three. For both strongly and weakly magnetic monomers we considered the influence of central attraction, ranging between 0 and 2$k_BT$. Magnetic field strength was chosen so that the complete range between the initial response and saturation was sampled for all types of MFs.

We find that for plain crosslinking even a slight increase of the central attraction leads to collapsed conformations for filaments and straightening field-dipole and dipole-dipole interactions, instead of unraveling the filament, optimise internal magnetisation to keep the maximum possible number of monomer-monomer contacts. As a result, with growing magnetic field, spherical symmetry of collapsed filament conformations transitions into cylindrical symmetry with the main axis aligned with the applied field. Even in the absence of central attraction, plainly crosslinked filaments never exhibit rod-like conformations. The magnetisation of plainly crosslinked filaments with weakly magnetic colloids can be described with the modified field theory, independently from the values of central attraction. In contrast, for filaments with strongly magnetic colloids, increasing central attraction results in significant hindrance of the magnetisation in low and intermediate fields. Interestingly, we find that plain crosslinking is able to introduce bistability of filament conformations for a certain ratio between dipolar and steric forces. For this narrow range of parameters, we observe two distinct conformations of filaments with well separated, distinct energies and entropy profiles. This might be an indication of a critical point, similarly to the case of Stockmayer filaments with ferromagnetic colloids. 

Behavior of filaments can be altered significantly by changing the crosslinking mechanism. In fact, additional directional correlations between colloids lead to more linear conformations and much stronger response to applied fields: both on a structural level and in terms of magnetisation. However, even in this case, MFs assume rod-like conformations only if the central attraction is weak enough. In fact, they tend to form structures in which each monomer has three neighbours, ever more so, as Van der Waals forces grow.

All this suggests that super-paramagnetic particles inside MFs behave similarly to ferromagnetic particles in Stockmayer fluids, in which, for the same range of parameters, compact droplet-like structures form. These structures deform into needles in strong applied fields, due to a vapour-liquid phase transition. However, the transition point in MFs with super-paramagnetic particles strongly depends on the colloids saturation magnetisation, applied magnetic field and crosslinking mechanism.

Currently, we are working on gaining a deeper insight into the phase behaviour of MFs of different length, with super-paramagnetic colloids and are investigating possibilities to control it by external mechanical, rheological and magnetic stimuli.

\section{Acknowledgements}
This research has been supported by the Russian Science Foundation Grant No.19-72-10033. S.S.K acknowledges support from the Austrian Science Fund (FWF), START-Projekt Y 627-N27.

\bibliography{magfils}

\begin{thebibliography}{10}
\expandafter\ifx\csname url\endcsname\relax
  \def\url#1{\texttt{#1}}\fi
\expandafter\ifx\csname urlprefix\endcsname\relax\def\urlprefix{URL }\fi
\expandafter\ifx\csname href\endcsname\relax
  \def\href#1#2{#2} \def\path#1{#1}\fi

\bibitem{resler1964magnetocaloric}
E.~Resler, R.~Rosensweig, Magnetocaloric power, AIAA Journal 2~(8) (1964)
  1418--1422.

\bibitem{2009-odenbach}
S.~Odenbach (Ed.), Colloidal Magnetic Fluids, Vol. 763 of Lecture Notes in
  Physics, Springer-Verlag, Berlin Heidelberg, 2009.
\newblock \href {http://dx.doi.org/10.1007/978-3-540-85387-9}
  {\path{doi:10.1007/978-3-540-85387-9}}.

\bibitem{zrinyi1998kinetics}
M.~Zr{\i}nyi, D.~Szab{\'o}, H.-G. Kilian, Kinetics of the shape change of
  magnetic field sensitive polymer gels, Polymer Gels and Networks 6~(6) (1998)
  441--454.

\bibitem{weeber2018polymer}
R.~Weeber, M.~Hermes, A.~M. Schmidt, C.~Holm, Polymer architecture of magnetic
  gels: a review., Journal of physics. Condensed matter: an Institute of
  Physics journal 30~(6) (2018) 063002--063002.

\bibitem{volkova2017motion}
T.~Volkova, V.~B{\"o}hm, T.~Kaufhold, J.~Popp, F.~Becker, D.~Y. Borin,
  G.~Stepanov, K.~Zimmermann, Motion behaviour of magneto-sensitive elastomers
  controlled by an external magnetic field for sensor applications, Journal of
  Magnetism and Magnetic Materials 431 (2017) 262--265.

\bibitem{frank1993voltage}
S.~Frank, P.~C. Lauterbur, Voltage-sensitive magnetic gels as magnetic
  resonance monitoring agents, Nature 363~(6427) (1993) 334.

\bibitem{Weeber_2012}
R.~Weeber, S.~Kantorovich, C.~Holm,
  \href{http://dx.doi.org/10.1039/C2SM26097B}{Deformation mechanisms in 2d
  magnetic gels studied by computer simulations}, Soft Matter 8~(38) (2012)
  9923.
\newblock \href {http://dx.doi.org/10.1039/c2sm26097b}
  {\path{doi:10.1039/c2sm26097b}}.
\newline\urlprefix\url{http://dx.doi.org/10.1039/C2SM26097B}

\bibitem{Dreyfus_2005}
R.~Dreyfus, J.~Baudry, M.~L. Roper, M.~Fermigier, H.~A. Stone, J.~Bibette,
  \href{http://dx.doi.org/10.1038/nature04090}{Microscopic artificial
  swimmers}, Nature 437~(7060) (2005) 862--865.
\newblock \href {http://dx.doi.org/10.1038/nature04090}
  {\path{doi:10.1038/nature04090}}.
\newline\urlprefix\url{http://dx.doi.org/10.1038/nature04090}

\bibitem{2008-benkoski}
J.~J. Benkoski, S.~E. Bowles, R.~L. Jones, J.~F. Douglas, J.~Pyun, A.~Karim,
  Self-assembly of polymer-coated ferromagnetic nanoparticles into mesoscopic
  polymer chains, J Polym Sci, Part B: Polym Phy 46~(20) (2008) 2267--2277.
\newblock \href {http://dx.doi.org/10.1002/polb.21558}
  {\path{doi:10.1002/polb.21558}}.

\bibitem{1998-furst}
E.~M. Furst, C.~Suzuki, M.~Fermigier, A.~P. Gast, Permanently linked
  monodisperse paramagnetic chains, Langmuir 14~(26) (1998) 7334--7336.
\newblock \href {http://dx.doi.org/10.1021/la980703i}
  {\path{doi:10.1021/la980703i}}.

\bibitem{1999-furst}
E.~M. Furst, A.~P. Gast, Micromechanics of dipolar chains using optical
  tweezers, Phys Rev Lett 82~(20) (1999) 4130--4133.
\newblock \href {http://dx.doi.org/10.1103/PhysRevLett.82.4130}
  {\path{doi:10.1103/PhysRevLett.82.4130}}.

\bibitem{WANG_2011}
H.~WANG, Y.~YU, Y.~SUN, Q.~CHEN,
  \href{http://dx.doi.org/10.1142/S1793292011002305}{Magnetic nanochains: A
  review}, Nano 06~(01) (2011) 1--17.
\newblock \href {http://dx.doi.org/10.1142/s1793292011002305}
  {\path{doi:10.1142/s1793292011002305}}.
\newline\urlprefix\url{http://dx.doi.org/10.1142/S1793292011002305}

\bibitem{wang2014multifunctional}
H.~Wang, A.~Mararenko, G.~Cao, Z.~Gai, K.~Hong, P.~Banerjee, S.~Zhou,
  Multifunctional 1d magnetic and fluorescent nanoparticle chains for enhanced
  mri, fluorescent cell imaging, and combined photothermal/chemotherapy, ACS
  applied materials \& interfaces 6~(17) (2014) 15309--15317.

\bibitem{cebers2016flexible}
A.~Cebers, K.~Erglis, Flexible magnetic filaments and their applications,
  Advanced Functional Materials 26~(22) (2016) 3783--3795.

\bibitem{cai2018fluidic}
G.~Cai, S.~Wang, L.~Zheng, J.~Lin, A fluidic device for immunomagnetic
  separation of foodborne bacteria using self-assembled magnetic nanoparticle
  chains, Micromachines 9~(12) (2018) 624.

\bibitem{2005-dreyfus}
R.~Dreyfus, J.~Baudry, M.~L. Roper, M.~Fermigier, H.~A. Stone, J.~Bibette,
  Microscopic artificial swimmers, Nature 437~(7060) (2005) 862--865.
\newblock \href {http://dx.doi.org/10.1038/nature04090}
  {\path{doi:10.1038/nature04090}}.

\bibitem{2008-erglis-mh}
K.~\={E}rglis, L.~Alberte, A.~C\={e}bers, Thermal fluctuations of non-motile
  magnetotactic bacteria in ac magnetic fields, Magnetohydrodynamics 44~(3)
  (2008) 223--236.

\bibitem{fayol2013use}
D.~Fayol, G.~Frasca, C.~Le~Visage, F.~Gazeau, N.~Luciani, C.~Wilhelm, Use of
  magnetic forces to promote stem cell aggregation during differentiation, and
  cartilage tissue modeling, Advanced Materials 25~(18) (2013) 2611--2616.

\bibitem{gerbal2015refined}
F.~Gerbal, Y.~Wang, F.~Lyonnet, J.-C. Bacri, T.~Hocquet, M.~Devaud, A refined
  theory of magnetoelastic buckling matches experiments with ferromagnetic and
  superparamagnetic rods, Proceedings of the National Academy of Sciences
  112~(23) (2015) 7135--7140.

\bibitem{evans2007magnetically}
B.~Evans, A.~Shields, R.~L. Carroll, S.~Washburn, M.~Falvo, R.~Superfine,
  Magnetically actuated nanorod arrays as biomimetic cilia, Nano letters 7~(5)
  (2007) 1428--1434.

\bibitem{erglis2011three}
K.~{\=E}rglis, R.~Livanovi{\v{c}}s, A.~C{\=e}bers, Three dimensional dynamics
  of ferromagnetic swimmer, Journal of Magnetism and Magnetic Materials
  323~(10) (2011) 1278--1282.

\bibitem{2003-goubault}
C.~Goubault, P.~Jop, M.~Fermigier, J.~Baudry, E.~Bertrand, J.~Bibette, Flexible
  magnetic filaments as micromechanical sensors, Phys. Rev. Lett. 91 (2003)
  260802.
\newblock \href {http://dx.doi.org/10.1103/PhysRevLett.91.260802}
  {\path{doi:10.1103/PhysRevLett.91.260802}}.

\bibitem{2005-cohen-tannoudji}
L.~Cohen-Tannoudji, E.~Bertrand, L.~Bressy, C.~Goubault, J.~Baudry, J.~Klein,
  J.-F. m.~c. Joanny, J.~Bibette, Polymer bridging probed by magnetic colloids,
  Phys. Rev. Lett. 94 (2005) 038301.
\newblock \href {http://dx.doi.org/10.1103/PhysRevLett.94.038301}
  {\path{doi:10.1103/PhysRevLett.94.038301}}.

\bibitem{2005-singh-lm}
H.~Singh, P.~E. Laibinis, T.~A. Hatton, Rigid, superparamagnetic chains of
  permanently linked beads coated with magnetic nanoparticles. synthesis and
  rotational dynamics under applied magnetic fields, Langmuir 21~(24) (2005)
  11500--11509.
\newblock \href {http://dx.doi.org/10.1021/la0517843}
  {\path{doi:10.1021/la0517843}}.

\bibitem{2005-singh-nl}
H.~Singh, P.~E. Laibinis, T.~A. Hatton, Synthesis of flexible magnetic
  nanowires of permanently linked core--shell magnetic beads tethered to a
  glass surface patterned by microcontact printing, Nano Lett 5~(11) (2005)
  2149--2154.
\newblock \href {http://dx.doi.org/10.1021/nl051537j}
  {\path{doi:10.1021/nl051537j}}.

\bibitem{2007-martinez-pedrero}
F.~Mart\'inez-Pedrero, M.~Tirado-Miranda, A.~Schmitt, J.~Callejas-Fern\'andez,
  Formation of magnetic filaments: A kinetic study, Phys Rev E 76~(1) (2007)
  011405.
\newblock \href {http://dx.doi.org/10.1103/PhysRevE.76.011405}
  {\path{doi:10.1103/PhysRevE.76.011405}}.

\bibitem{2007-evans}
B.~A. Evans, A.~R. Shields, R.~L. Carroll, S.~Washburn, M.~R. Falvo,
  R.~Superfine, Magnetically actuated nanorod arrays as biomimetic cilia, Nano
  Lett 7~(5) (2007) 1428--1434.
\newblock \href {http://dx.doi.org/10.1021/nl070190c}
  {\path{doi:10.1021/nl070190c}}.

\bibitem{2009-zhou}
Z.~Zhou, G.~Liu, D.~Han, Coating and structural locking of dipolar chains of
  cobalt nanoparticles, ACS Nano 3~(1) (2009) 165--172.
\newblock \href {http://dx.doi.org/10.1021/nn8005366}
  {\path{doi:10.1021/nn8005366}}.

\bibitem{2011-benkoski}
J.~J. Benkoski, J.~L. Breidenich, O.~M. Uy, A.~T. Hayes, R.~M. Deacon, H.~B.
  Land, J.~M. Spicer, P.~Y. Keng, J.~Pyun, Dipolar organization and magnetic
  actuation of flagella-like nanoparticle assemblies, J Mater Chem 21 (2011)
  7314--7325.
\newblock \href {http://dx.doi.org/10.1039/C0JM04014B}
  {\path{doi:10.1039/C0JM04014B}}.

\bibitem{2011-wang}
H.~Wang, Y.~Yu, Y.~Sun, Q.~Chen, Magnetic nanochains: a review, Nano 06~(01)
  (2011) 1--17.
\newblock \href {http://dx.doi.org/10.1142/S1793292011002305}
  {\path{doi:10.1142/S1793292011002305}}.

\bibitem{2012-sarkar}
D.~Sarkar, M.~Mandal, Static and dynamic magnetic characterization of
  {DNA}-templated chain-like magnetite nanoparticles, The Journal of Physical
  Chemistry C 116~(5) (2012) 3227--3234.
\newblock \href {http://dx.doi.org/10.1021/jp208020z}
  {\path{doi:10.1021/jp208020z}}.

\bibitem{2012-breidenich}
J.~L. Breidenich, M.~C. Wei, G.~V. Clatterbaugh, J.~J. Benkoski, P.~Y. Keng,
  J.~Pyun, Controlling length and areal density of artificial cilia through the
  dipolar assembly of ferromagnetic nanoparticles, Soft Matter 8 (2012)
  5334--5341.
\newblock \href {http://dx.doi.org/10.1039/C2SM25096A}
  {\path{doi:10.1039/C2SM25096A}}.

\bibitem{busseron2013supramolecular}
E.~Busseron, Y.~Ruff, E.~Moulin, N.~Giuseppone, Supramolecular self-assemblies
  as functional nanomaterials, Nanoscale 5~(16) (2013) 7098--7140.

\bibitem{2014-byrom}
J.~Byrom, P.~Han, M.~Savory, S.~L. Biswal, Directing assembly of {DNA}-coated
  colloids with magnetic fields to generate rigid, semiflexible, and flexible
  chains, Langmuir 30~(30) (2014) 9045--9052.
\newblock \href {http://dx.doi.org/10.1021/la5009939}
  {\path{doi:10.1021/la5009939}}.

\bibitem{hill2014colloidal}
L.~J. Hill, J.~Pyun, Colloidal polymers via dipolar assembly of magnetic
  nanoparticle monomers, ACS applied materials \& interfaces 6~(9) (2014)
  6022--6032.

\bibitem{2015-bannwarth}
M.~B. Bannwarth, S.~Utech, S.~Ebert, D.~A. Weitz, D.~Crespy, K.~Landfester,
  Colloidal polymers with controlled sequence and branching constructed from
  magnetic field assembled nanoparticles, ACS Nano 9 (2015) 2720--2728.
\newblock \href {http://dx.doi.org/10.1021/nn5065327}
  {\path{doi:10.1021/nn5065327}}.

\bibitem{Maye_2009}
M.~M. Maye, D.~Nykypanchuk, M.~Cuisinier, D.~van~der Lelie, O.~Gang,
  \href{http://dx.doi.org/10.1038/nmat2421}{Stepwise surface encoding for
  high-throughput assembly of nanoclusters}, Nature Materials 8~(5) (2009)
  388--391.
\newblock \href {http://dx.doi.org/10.1038/nmat2421}
  {\path{doi:10.1038/nmat2421}}.
\newline\urlprefix\url{http://dx.doi.org/10.1038/nmat2421}

\bibitem{Sun_2012}
D.~Sun, A.~L. Stadler, M.~Gurevich, E.~Palma, E.~Stach, D.~van~der Lelie,
  O.~Gang, \href{http://dx.doi.org/10.1039/c2nr31908j}{Heterogeneous
  nanoclusters assembled by pna-templated double-stranded dna}, Nanoscale
  4~(21) (2012) 6722.
\newblock \href {http://dx.doi.org/10.1039/c2nr31908j}
  {\path{doi:10.1039/c2nr31908j}}.
\newline\urlprefix\url{http://dx.doi.org/10.1039/c2nr31908j}

\bibitem{Sun_2013}
D.~Sun, O.~Gang, \href{http://dx.doi.org/10.1021/la4000186}{Dna-functionalized
  quantum dots: Fabrication, structural, and physicochemical properties},
  Langmuir 29~(23) (2013) 7038--7046.
\newblock \href {http://dx.doi.org/10.1021/la4000186}
  {\path{doi:10.1021/la4000186}}.
\newline\urlprefix\url{http://dx.doi.org/10.1021/la4000186}

\bibitem{Zhang_2013}
Y.~Zhang, F.~Lu, K.~G. Yager, D.~van~der Lelie, O.~Gang,
  \href{http://dx.doi.org/10.1038/nnano.2013.209}{A general strategy for the
  dna-mediated self-assembly of functional nanoparticles into heterogeneous
  systems}, Nature Nanotechnology 8~(11) (2013) 865--872.
\newblock \href {http://dx.doi.org/10.1038/nnano.2013.209}
  {\path{doi:10.1038/nnano.2013.209}}.
\newline\urlprefix\url{http://dx.doi.org/10.1038/nnano.2013.209}

\bibitem{Srivastava_2014}
S.~Srivastava, D.~Nykypanchuk, M.~Fukuto, J.~D. Halverson, A.~V. Tkachenko,
  K.~G. Yager, O.~Gang,
  \href{http://dx.doi.org/10.1021/ja501749b}{Two-dimensional dna-programmable
  assembly of nanoparticles at liquid interfaces}, Journal of the American
  Chemical Society 136~(23) (2014) 8323--8332.
\newblock \href {http://dx.doi.org/10.1021/ja501749b}
  {\path{doi:10.1021/ja501749b}}.
\newline\urlprefix\url{http://dx.doi.org/10.1021/ja501749b}

\bibitem{Tian_2015}
Y.~Tian, T.~Wang, W.~Liu, H.~L. Xin, H.~Li, Y.~Ke, W.~M. Shih, O.~Gang,
  \href{http://dx.doi.org/10.1038/nnano.2015.105}{Prescribed nanoparticle
  cluster architectures and low-dimensional arrays built using octahedral dna
  origami frames}, Nature Nanotechnology 10~(7) (2015) 637--644.
\newblock \href {http://dx.doi.org/10.1038/nnano.2015.105}
  {\path{doi:10.1038/nnano.2015.105}}.
\newline\urlprefix\url{http://dx.doi.org/10.1038/nnano.2015.105}

\bibitem{liu2016self}
W.~Liu, J.~Halverson, Y.~Tian, A.~V. Tkachenko, O.~Gang, Self-organized
  architectures from assorted dna-framed nanoparticles, Nature chemistry 8~(9)
  (2016) 867.

\bibitem{2003-cebers}
A.~C\=ebers, Dynamics of a chain of magnetic particles connected with elastic
  linkers, J. Phys.: Condens. Matter 15~(15) (2003) S1335.
\newblock \href {http://dx.doi.org/10.1088/0953-8984/15/15/303}
  {\path{doi:10.1088/0953-8984/15/15/303}}.

\bibitem{2004-shcherbakov}
V.~P. Shcherbakov, M.~Winklhofer, Bending of magnetic filaments under a
  magnetic field, Phys. Rev. E 70 (2004) 061803.
\newblock \href {http://dx.doi.org/10.1103/PhysRevE.70.061803}
  {\path{doi:10.1103/PhysRevE.70.061803}}.

\bibitem{2004-cebers}
A.~C\=ebers, I.~Javaitis, Dynamics of a flexible magnetic chain in a rotating
  magnetic field, Phys Rev E 69~(2) (2004) 021404.
\newblock \href {http://dx.doi.org/10.1103/PhysRevE.69.021404}
  {\path{doi:10.1103/PhysRevE.69.021404}}.

\bibitem{2005-cebers}
A.~C\={e}bers, Flexible magnetic filaments, Curr Opin Colloid Interface Sci
  10~(3-4) (2005) 167--175.
\newblock \href {http://dx.doi.org/10.1016/j.cocis.2005.07.002}
  {\path{doi:10.1016/j.cocis.2005.07.002}}.

\bibitem{belovs2006nonlinear}
M.~Belovs, A.~C{\=e}bers, Nonlinear dynamics of semiflexible magnetic filaments
  in an ac magnetic field, Physical Review E 73~(5) (2006) 051503.

\bibitem{cebers2007magnetic}
A.~C{\=e}bers, T.~C{\=\i}rulis, Magnetic elastica, Physical Review E 76~(3)
  (2007) 031504.

\bibitem{_rglis_2008}
K.~{\=E}rglis, D.~Zhulenkovs, A.~Sharipo, A.~C{\=e}bers,
  \href{http://dx.doi.org/10.1088/0953-8984/20/20/204107}{Elastic properties of
  dna linked flexible magnetic filaments}, Journal of Physics: Condensed Matter
  20~(20) (2008) 204107.
\newblock \href {http://dx.doi.org/10.1088/0953-8984/20/20/204107}
  {\path{doi:10.1088/0953-8984/20/20/204107}}.
\newline\urlprefix\url{http://dx.doi.org/10.1088/0953-8984/20/20/204107}

\bibitem{kuznetsov2019equilibrium}
A.~A. Kuznetsov, Equilibrium properties of magnetic filament suspensions,
  Journal of Magnetism and Magnetic Materials 470 (2019) 28--32.

\bibitem{huang2016buckling}
S.~Huang, G.~Pessot, P.~Cremer, R.~Weeber, C.~Holm, J.~Nowak, S.~Odenbach,
  A.~M. Menzel, G.~K. Auernhammer, Buckling of paramagnetic chains in soft
  gels, Soft Matter 12~(1) (2016) 228--237.

\bibitem{zhao2018nonlinear}
J.~Zhao, D.~Du, S.~L. Biswal, Nonlinear multimode buckling dynamics examined
  with semiflexible paramagnetic filaments, Physical Review E 98~(1) (2018)
  012602.

\bibitem{wei2016assembly}
J.~Wei, F.~Song, J.~Dobnikar, Assembly of superparamagnetic filaments in
  external field, Langmuir 32~(36) (2016) 9321--9328.

\bibitem{kuei2017strings}
S.~Kuei, B.~Garza, S.~L. Biswal, From strings to coils: Rotational dynamics of
  dna-linked colloidal chains, Physical Review Fluids 2~(10) (2017) 104102.

\bibitem{dempster2017contractile}
J.~M. Dempster, P.~V{\'a}zquez-Montejo, M.~O. de~la Cruz, Contractile actuation
  and dynamical gel assembly of paramagnetic filaments in fast precessing
  fields, Physical Review E 95~(5) (2017) 052606.

\bibitem{vazquez2017paramagnetic}
P.~V{\'a}zquez-Montejo, J.~M. Dempster, M.~O. de~la Cruz, Paramagnetic
  filaments in a fast precessing field: Planar versus helical conformations,
  Physical Review Materials 1~(6) (2017) 064402.

\bibitem{2006-gauger}
E.~Gauger, H.~Stark, Numerical study of a microscopic artificial swimmer, Phys.
  Rev. E 74 (2006) 021907.
\newblock \href {http://dx.doi.org/10.1103/PhysRevE.74.021907}
  {\path{doi:10.1103/PhysRevE.74.021907}}.

\bibitem{roper2006dynamics}
M.~Roper, R.~Dreyfus, J.~Baudry, M.~Fermigier, J.~Bibette, H.~A. Stone, On the
  dynamics of magnetically driven elastic filaments, Journal of Fluid Mechanics
  554 (2006) 167--190.

\bibitem{roper2008magnetic}
M.~Roper, R.~Dreyfus, J.~Baudry, M.~Fermigier, J.~Bibette, H.~A. Stone, Do
  magnetic micro-swimmers move like eukaryotic cells?, Proceedings of the Royal
  Society A: Mathematical, Physical and Engineering Sciences 464~(2092) (2008)
  877--904.

\bibitem{philippova2011magnetic}
O.~Philippova, A.~Barabanova, V.~Molchanov, A.~Khokhlov, Magnetic polymer
  beads: Recent trends and developments in synthetic design and applications,
  European polymer journal 47~(4) (2011) 542--559.

\bibitem{pak2011high}
O.~S. Pak, W.~Gao, J.~Wang, E.~Lauga, High-speed propulsion of flexible
  nanowire motors: Theory and experiments, Soft Matter 7~(18) (2011)
  8169--8181.

\bibitem{Saanchez_2015}
P.~A. S{\'a}nchez, E.~S. Pyanzina, E.~V. Novak, J.~J. Cerd{\`a}, T.~Sintes,
  S.~S. Kantorovich,
  \href{http://dx.doi.org/10.1021/acs.macromol.5b01086}{Supramolecular magnetic
  brushes: The impact of dipolar interactions on the equilibrium structure},
  Macromolecules 48~(20) (2015) 7658--7669.
\newblock \href {http://dx.doi.org/10.1021/acs.macromol.5b01086}
  {\path{doi:10.1021/acs.macromol.5b01086}}.
\newline\urlprefix\url{http://dx.doi.org/10.1021/acs.macromol.5b01086}

\bibitem{biswal2004micromixing}
S.~L. Biswal, A.~P. Gast, Micromixing with linked chains of paramagnetic
  particles, Analytical chemistry 76~(21) (2004) 6448--6455.

\bibitem{yang2017magnetic}
T.~Yang, T.~O. Tasci, K.~B. Neeves, N.~Wu, D.~W. Marr, Magnetic microlassos for
  reversible cargo capture, transport, and release, Langmuir 33~(23) (2017)
  5932--5937.

\bibitem{lopez2009magnetorheology}
M.~T. L{\'o}pez-L{\'o}pez, P.~Kuzhir, G.~Bossis, Magnetorheology of fiber
  suspensions. i. experimental, Journal of Rheology 53~(1) (2009) 115--126.

\bibitem{weeks1971role}
J.~D. Weeks, D.~Chandler, H.~C. Andersen, Role of repulsive forces in
  determining the equilibrium structure of simple liquids, The Journal of
  chemical physics 54~(12) (1971) 5237--5247.

\bibitem{ivanov2001magnetic}
A.~O. Ivanov, O.~B. Kuznetsova, Magnetic properties of dense ferrofluids: An
  influence of interparticle correlations, Physical Review E 64~(4) (2001)
  041405.

\bibitem{elfimova2019static}
E.~A. Elfimova, A.~O. Ivanov, P.~J. Camp, Static magnetization of
  immobilized{,} weakly interacting{,} superparamagnetic nanoparticles,
  Nanoscale (2019) --\href {http://dx.doi.org/10.1039/C9NR07425B}
  {\path{doi:10.1039/C9NR07425B}}.

\bibitem{2006-limbach}
H.~J. Limbach, A.~Arnold, B.~A. Mann, C.~Holm, {ESPResSo} -- an extensible
  simulation package for research on soft matter systems, Comput Phys Commun
  174~(9) (2006) 704--727.
\newblock \href {http://dx.doi.org/10.1016/j.cpc.2005.10.005}
  {\path{doi:10.1016/j.cpc.2005.10.005}}.

\bibitem{allen2017computer}
M.~P. Allen, D.~J. Tildesley, Computer simulation of liquids, Oxford university
  press, 2017.

\bibitem{2007-ivanov}
V.~A. Ivanov, J.~A. Martemyanova, Monte carlo computer simulation of a single
  semi-flexible macromolecule at a plane surface, Macromol Symp 252~(1) (2007)
  12--23.
\newblock \href {http://dx.doi.org/10.1002/masy.200750602}
  {\path{doi:10.1002/masy.200750602}}.

\bibitem{moore2015liquid}
S.~G. Moore, M.~J. Stevens, G.~S. Grest, Liquid-vapor interface of the
  stockmayer fluid in a uniform external field, Physical Review E 91~(2) (2015)
  022309.

\bibitem{van1993makes}
M.~Van~Leeuwen, B.~Smit, What makes a polar liquid a liquid?, Physical review
  letters 71~(24) (1993) 3991.

\bibitem{panagiotopoulos1992direct}
A.~Z. Panagiotopoulos, Direct determination of fluid phase equilibria by
  simulation in the gibbs ensemble: a review, Molecular simulation 9~(1) (1992)
  1--23.

\bibitem{stevens1995structure}
M.~J. Stevens, G.~S. Grest, Structure of soft-sphere dipolar fluids, Physical
  Review E 51~(6) (1995) 5962.

\bibitem{adams1981static}
D.~Adams, E.~Adams, Static dielectric properties of the stockmayer fluid from
  computer simulation, Molecular Physics 42~(4) (1981) 907--926.

\bibitem{novak2019structure}
E.~V. Novak, E.~S. Pyanzina, P.~A. S{\'a}nchez, S.~S. Kantorovich, The
  structure of clusters formed by stockmayer supracolloidal magnetic polymers,
  The European Physical Journal E 42~(12) (2019) 158.

\bibitem{2016-cerda-pccp}
J.~J. Cerd\`{a}, P.~A. S\'{a}nchez, D.~L\"{u}sebrink, S.~S. Kantorovich,
  T.~Sintes, Flexible magnetic filaments under the influence of external
  magnetic fields in the limit of infinite dilution, Phys. Chem. Chem. Phys. 18
  (2016) 12616--12625.
\newblock \href {http://dx.doi.org/10.1039/C6CP00923A}
  {\path{doi:10.1039/C6CP00923A}}.

\end{thebibliography}
\bibliographystyle{elsarticle-num}

% biblio

\end{document}